\documentclass[traditabstract,onecolumn]{aa}
\idline{A\&A 566, A8 (2014)}{1}%
\makeatletter
\def\@doi{\href{http://doi.org/10.1051/0004-6361/201220021}{DOI: 10.1051/0004-6361/201220021}}
\makeatother
\usepackage{amsmath}
\usepackage{natbib}
\bibpunct{(}{)}{;}{a}{}{,}
\usepackage{textcomp}
\usepackage[T1]{fontenc}
\usepackage{txfonts}
\usepackage{graphicx}
\usepackage{hyperref}
\newcommand\Aspects{\textsc{Aspects}}
\newcommand\Iras{\textsc{Iras}}
\newcommand\card{\mathop{\#}\mathopen{}} 
\newcommand\transpose[1]{#1^{\textsf{\textsc{t}}}} 
\newcommand\df{\textnormal{d}} 
\newcommand\radian{\textnormal{rad}}
\newcommand\comma{{,}\,}
\newcommand\multspace{\mathclose{}\,\mathopen{}} 
\newcommand\integinterv[2]{\llbracket#1{,}\,\mathopen{}#2\rrbracket} 
\newcommand\Left{\mathopen{}\mathclose\bgroup\left}
\newcommand\Right{\aftergroup\egroup\right}
\newcommand\leftsubstack[1]{\subarray{l}#1\endsubarray}
\DeclareMathOperator{\Rot}{Rot}
\DeclareMathOperator{\Diag}{Diag}
\newcommand\sto{{\textnormal{s:o}}} 
\newcommand\ots{{\textnormal{o:s}}}
\newcommand\oto{{\textnormal{o:o}}} 
\newcommand\loc{{\textnormal{loc}}} 
\newcommand\ato{{\textnormal{:o}}} 
\newcommand\wrong{{\textnormal{w}}} 
\newcommand\Lh{L}
\newcommand\Lhsto{\Lh_\sto}
\newcommand\Lhoto{\Lh_\oto}
\newcommand\Lhots{\Lh_\ots}
\newcommand\Prob{P}
\newcommand\Ploc{\Prob_{\!\loc}}
\newcommand\Psto{\Prob_{\!\sto}}
\newcommand\Poto{\Prob_{\!\oto}}
\newcommand\Pato{\Prob_{\!\ato}}
\newcommand\sigmatot{\mathring\sigma}
\newcommand\nutot{\mathring\nu}
\newcommand\uri{\vec u_{r\comma i}}
\newcommand\udi{\vec u_{\delta\comma i}}
\newcommand\uai{\vec u_{\alpha\comma i}}
\newcommand\urpj{\vec u_{r'\!\comma j}}
\newcommand\udpj{\vec u_{\delta'\!\comma j}}
\newcommand\uapj{\vec u_{\alpha'\!\comma j}}
\newcommand\deltapj{\delta'_{\smash[t]{j}}}
\newcommand\alphapj{\alpha'_{\smash[t]{j}}}
\newcommand\vrpj{\vec r'_{\smash[t]{j}}}
\newcommand\vrzi{\vec r_{0\comma i}}
\newcommand\vrpzj{\vec r'_{\smash[t]{0\comma j}}}
\newcommand\betapj{\beta'_{\smash[t]{j}}}
\newcommand\gammapj{\gamma'_{\smash[t]{j}}}
\newcommand\thetapj{\theta'_{\smash[t]{j}}}
\newcommand\Gammapj{\Gamma'_{\smash[t]{j}}}
\newcommand\Gammazi{\Gamma_{0\comma i}}
\newcommand\Stot{S\mkern-2mu}
\newcommand\Si{S\mkern-2mu{}_i}
\newcommand\Npi{N'_{\smash[t]{i}}}
\newcommand\COORDpi{C'_{\smash[t]{i}}}
\newcommand\coordpj{c'_{\smash[t]{j}}}
\providecommand\cramped[1]{{\kern-\nulldelimiterspace\radical0{#1}}}
\newcommand\np{{\cramped{n'}}}
\newcommand\npeff{{\cramped{n'_{\smash[t]{\textnormal{eff}}}}}}
\newcommand\npi{\cramped{n'_{\smash[t]{i}}}}
\newcommand\Mp{M'\futurelet\next\MpI}
\newcommand\MpI{\if\next_\expandafter\MpII\fi}
\def\MpII_#1{_{\smash[t]{#1}}}
\newcommand\rhopi{\rho'_{\smash[t]{i}}}
\newcommand\jMLC{j_{\textnormal{MLC}}}
\newcommand\depthmin{d_{\textnormal{min}}}
\begin{document}
\setlength{\abovedisplayskip}{\belowdisplayshortskip}
\setlength{\belowdisplayskip}{\belowdisplayshortskip}
\title{%
  Probabilistic positional association of catalogs\\
  of astrophysical sources: the \Aspects\ code%
  \thanks{Available at \href{http://www2.iap.fr/users/fioc/Aspects/}{www2.iap.fr/users/fioc/Aspects/}.}
}
\author{Michel Fioc}
\institute{%
  Institut d'astrophysique de Paris, UPMC~-~univ.~Paris~6, CNRS,
  UMR~7095, 98\emph{bis}~boulevard Arago, \mbox{F-75014} Paris, France;\\
  \email{Michel.Fioc@iap.fr}
}
\date{Received 16 July 2012 / Accepted 1 November 2013}
\abstract{
  We describe a probabilistic method of cross-identifying astrophysical
  sources in two catalogs from their positions and positional uncertainties.
  The probability that an object is associated with a source from the other 
  catalog, or that it has no counterpart, is derived under two exclusive 
  assumptions:
  first, the classical case of several-to-one associations, and then 
  the more realistic but more difficult problem of one-to-one associations.

  In either case, the likelihood of observing the objects in the two catalogs 
  at their effective positions is computed and a maximum likelihood estimator 
  of the fraction of sources with a counterpart --~a quantity needed 
  to compute the probabilities of association~-- is built.
  When the positional uncertainty in one or both catalogs is unknown, 
  this method may be used to estimate its typical value and even to study 
  its dependence on the size of objects.
  It may also be applied when the true centers of a source 
  and of its counterpart at another wavelength do not coincide.

  To compute the likelihood and association probabilities under the different 
  assumptions, we developed a Fortran~95 code called \Aspects\
  ([asp$\varepsilonup$],
  ``\!\emph{\textsc{\textbf{As}}sociation
    \textsc{\textbf{p}}ositionnell\textsc{\textbf{e}}\slash
    \textsc{\textbf{p}}robabilist\textsc{\textbf{e}}
    de \textsc{\textbf{c}}a\textsc{\textbf{t}}alogues
    de \textsc{\textbf{s}}ources}''
  in French);
  its source files are made freely available.
  To test \Aspects, all-sky mock catalogs containing up to $10^5$ objects 
  were created, forcing either several-to-one or one-to-one associations.
  The analysis of these simulations confirms that, in both cases, 
  the assumption with the highest likelihood is the right one and
  that estimators of unknown parameters built for the appropriate association 
  model are reliable.
}
\titlerunning{%
  Probabilistic positional association of catalogs of astrophysical sources: 
  the \Aspects\ code
}
\authorrunning{M. Fioc}
\maketitle
\section{Introduction}
The most basic method of cross-identifying two catalogs
$K$ and $K'$ with known circular positional uncertainties
is to consider that a $K'$-source $\Mp$ is the same as an object $M$ of $K$
if it falls within a disk centered on $M$ and having a radius equal to
a few times their combined positional uncertainty;
if the disk is void, $M$ has no counterpart,
and if it contains several $K'$-sources, the nearest one
is identified to $M$.
This solution is defective for several reasons:
it does not take the density of sources into account;
positional uncertainty ellipses are not properly treated;
the radius of the disk is arbitrary;
positional uncertainties are not always known;
$K$ and $K'$ do not play symmetrical roles;
the identification is ambiguous if a $K'$-source may be associated to several
objects of $K$.
Worst of all, it does not provide a \emph{probability} of association.

Beyond this na\"\i{}ve method, the cross-identification problem has been 
studied by \citet{Condon}, \citet{DeRuiter}, \citet{Prestage}, \citet{SS},
\citet{Bauer}, and \citet{Rutledge}, among others.
As shown by the recent papers of \citet{BS}, \citet{Brand}, \citet{Rohde}, 
\citet{Roseboom}, and \citet{Pineau}, this field is still very active and 
will be more so with the wealth of forthcoming multiwavelength data
and the virtual observatory \citep{Vignali}.
In these papers, the identification is performed using a ``likelihood ratio''.
For two objects $(M, \Mp) \in K\times K'$ with known coordinates and
positional uncertainties, and given the local surface density of $K'$-sources,
this ratio is typically computed as
\begin{equation}
  \label{def_LR}
  \lambda \coloneqq
  \frac{
    \Prob(\text{position} \mid \text{counterpart})
  }{
    \Prob(\text{position} \mid \text{chance})
  },
\end{equation}
where $\Prob(\text{position} \mid \text{counterpart})$ is the probability 
of finding $\Mp$ at some position relative to $M$ if $\Mp$ is a counterpart 
of $M$, and $\Prob(\text{position} \mid \text{chance})$ is the probability 
that $\Mp$ is there by chance.
As noticed by \citet{SS}, there has been some confusion when defining
and interpreting $\lambda$, and, more importantly, in deriving
the probability%
\footnote{%
  For instance, \citet{DeRuiter}
  wrongly
  state that, if there is
  a counterpart, the closest object is always the right one.
}
that $M$ and $\Mp$ are the same.

To associate sources from catalogs at different wavelengths,
some authors include some \emph{a priori}
information on the spectral energy distribution (SED) of the objects
in this likelihood ratio.
When this work started, our primary goal was to build template observational
SEDs from the optical to the far-infrared for different types of galaxies.
We initially intended to cross-identify the \Iras\ Faint Source Survey
\citep{FSS, FSC} with the \textsc{Leda} database \citep{LEDA}.
Because of the high positional inaccuracy of \Iras\ data,
special care was needed to identify optical sources with infrared ones.
While \Iras\ data are by now quite outdated and have been superseded by
\emph{Spitzer} and \emph{Herschel} observations, we think that the procedure we
began to develop at that time may be valuable for other studies.
Because we aimed to fit synthetic SEDs to the template observational
ones, we could not and did not want to make assumptions on the SED of
objects based on their type, since this would have biased the procedure.
We therefore rely only on positions in what follows.

The method we use is in essence similar to that of \citet{SS}.
Because thinking in terms of probabilities rather than of
likelihood ratios highlights some implicit assumptions, we found it however
useful for the sake of clarity to detail hereafter our calculations.
This allows us moreover to propose a systematic way to estimate 
the unknown parameters required to compute the probabilities of association 
and to extend our work to a case not covered by the papers cited above 
(see Sect.~\ref{one}).

After some preliminaries (Sect.~\ref{notat_assump}),
we compute in Sect.~\ref{several} the probability of association
under the hypothesis that a $K$-source has at most one counterpart
in $K'$ but that several objects of $K$ may share
the same one (``several-to-one'' associations).
We also compute the likelihood to observe all the sources at their effective 
positions and use it to estimate the fraction of objects with a counterpart 
and, if unknown, the positional uncertainty in one or both catalogs.
In Sect.~\ref{one}, we do the same calculations under the assumption
that a $K$-source has at most one counterpart in $K'$ and that no other
object of $K$ has the same counterpart (``one-to-one'' associations).
In Sect.~\ref{code}, we present a code, \Aspects, implementing the results
of Sects.~\ref{several} and \ref{one}, and with which we compute 
the likelihoods and probabilities of association under the aforementioned
assumptions.
We test it on simulations in Sect.~\ref{simul}.
The probability distribution of the relative positions
of associated sources is modeled in App.~\ref{cov}.
\section{Preliminaries}
\label{notat_assump}
\subsection{Notations}
\label{notations}
We consider two catalogs $K$ and $K'$ defined on a common surface of the sky, 
of area $\Stot$, and containing respectively $n$ sources
$(M_i)_{i\in\integinterv{1}{n}}$ and $\np$ sources
$(\Mp_j)^{}_{\smash[t]{j\in\integinterv{1}{\np}}}$.
We define the following events:
\begin{itemize}
\item
  $c_i$: $M_i$ is in the infinitesimal surface element $\df^2\vec r_i$ located 
  at $\vec r_i$;
\item
  $\coordpj$: $\Mp_j$ is in the infinitesimal surface element $\df^2\vrpj$ 
  located at $\vrpj$;
\item
  $C \coloneqq \bigcap_{i=1}^n c_i$: the coordinates of all $K$-sources 
  are known;
\item
  $C' \coloneqq \bigcap_{j=1}^\np \coordpj$: the coordinates of all $K'$-sources 
  are known;
\item
  $A_{i\comma j}$, with $i \neq 0$ and $j \neq 0$: $\Mp_j$ is a counterpart 
  of $M_i$;\vadjust{\vspace{-1pt}}
\item
  $A_{i\comma 0}$: $M_i$ has no counterpart in $K'$, i.e.\ 
  $A_{i\comma 0} = \overline{\bigcup_{j\neq0}A_{i\comma j}}$,
  where $\overline{\omega}$ is the negation of an event $\omega$;
\item
  $A_{0\comma j}$: $\Mp_j$ has no counterpart in $K$.
\end{itemize}

We denote by $f$ (resp.~$f'$) the unknown \emph{a priori}
(i.e., not knowing the coordinates) probability that any
element of $K$ (resp.~$K'$) has a counterpart in $K'$ (resp.~$K$).
In terms of the events $(A_{i\comma j})$, for any $(M_i, \Mp_j) \in K \times K'$,
\begin{equation}
  \label{def_f}
  \Prob\Bigl(\bigcup_{k\neq0} A_{i\comma k}\Bigr) = f;
  \qquad
  \Prob(A_{i\comma 0}) = 1-f;
  \qquad
  \Prob\Bigl(\bigcup_{k\neq0} A_{k\comma j}\Bigr) = f';
  \qquad
  \Prob(A_{0\comma j}) = 1-f'.
\end{equation}
We see in Sects.~\ref{fractionsto} and \ref{fractionoto} how 
to estimate $f$ and $f'$.

The angular distance between two points $Y$ and $Z$ is written
$\psi(Y, Z)$.
More specifically, we put $\psi_{i\comma j} = \psi(M_i, \Mp_j)$.
\subsection{Assumptions}
\label{assump}
Calculations are carried out under one of three exclusive assumptions:
\begin{itemize}
\item
  \emph{Several-to-one hypothesis}:
  \begin{equation}
    \Left\{
    \begin{aligned}
      &\text{for all $M_i$, the events $(A_{i\comma j})_{j\in\integinterv{1}{\np}}$ 
        are exclusive};
      \\
      &\text{for all $\Mp_j$, the events $(A_{i\comma j})_{i\in\integinterv{1}{n}}$ 
        are independent}.
    \end{aligned}
    \Right.
    \tag{$H_\sto$}
    \label{H_sto}
  \end{equation}
  Therefore, a $K$-source has at most one counterpart in $K'$,
  but a $K'$-source may have several counterparts in $K$.
  Since more $K$-sources have a counterpart in $K'$ than the converse,
  $f\multspace n \geqslant f'\multspace \np$.
  This assumption is reasonable if the angular resolution in $K'$ (e.g.\ \Iras)
  is much poorer than in $K$ (e.g.\ \textsc{Leda}), since several distinct 
  objects of $K$ may then be confused in $K'$.
\item
  \emph{One-to-several hypothesis}:
  the symmetric of assumption~\ref{H_sto}, i.e.,
  \begin{equation}
    \Left\{
    \begin{aligned}
      &\text{for all $M_i$, the events $(A_{i\comma j})_{i\in\integinterv{1}{n}}$
        are independent};
      \\
      &\text{for all $\Mp_j$, the events $(A_{i\comma j})_{j\in\integinterv{1}{\np}}$
        are exclusive}.
    \end{aligned}
    \Right.
    \tag{$H_\ots$}
    \label{H_ots}
  \end{equation}
  In that case, $f\multspace n \leqslant f'\multspace \np$.
  This assumption is appropriate for catalogs of extended sources
  that, although observed as single at the wavelength of $K$,
  may look broken up at the wavelength of $K'$.
\item
  \emph{One-to-one hypothesis}:
  any $K$-source has at most one counterpart in $K'$ and reciprocally, i.e.
  \begin{equation}
    \text{all the events 
      $(A_{i\comma j})_{i\in\integinterv{1}{n}\comma j\in\integinterv{1}{\np}}$
      are exclusive}.
    \tag{$H_\oto$}
    \label{H_oto}
  \end{equation}
  Then, $f\multspace n = f'\multspace \np$.
  This assumption is the most relevant one for high-resolution catalogs of
  point sources or of well-defined extended sources.
\end{itemize}
Probabilities, likelihoods, and estimators specifically derived under either
assumption~\ref{H_sto}, \ref{H_ots}, or \ref{H_oto}
are written with the subscript ``\sto'', ``\ots'', or ``\oto'',
respectively;
the subscript ``\ato'' is used for results valid for both~\ref{H_sto}
and \ref{H_oto}.
The ``several-to-several'' hypothesis where all the events
$(A_{i\comma j})_{i\in\integinterv{1}{n}\comma j\in\integinterv{1}{\np}}$ are independent
is not considered here.

We make two other assumptions:
all the associations $A_{i\comma j}$ with $i \neq 0$ and $j \neq 0$
are considered \emph{a priori} as equally likely,
and the effect of clustering is negligible.
\subsection{Approach}
Our approach is the following.
For each of the assumptions \ref{H_sto}, \ref{H_oto}, and \ref{H_ots}, we
\begin{itemize}
\item
  find an expression for the probabilities of association,
\item
  build estimators of the unknown parameters needed to compute these
  probabilities, and
\item
  compute the likelihood of the assumption from the data.
\end{itemize}
Then, we compute the probabilities of association for the best estimators
of unknown parameters and the most likely assumption.

Although \ref{H_sto} is less symmetrical and neutral than
\ref{H_oto}, we begin our study with this assumption:
first, because computations are much simpler under \ref{H_sto}
than under \ref{H_oto} and serve as a guide for the latter;
second, because they provide initial values for the iterative procedure
(Sect.~\ref{working}) used to effectively compute probabilities under
\ref{H_oto}.
\section{Several-to-one associations}
\label{several}
In this section, we assume that hypothesis~\ref{H_sto} holds.
As shown in Sect.~\ref{local}, this is also the assumption implicitly 
made by the authors cited in the introduction.
\subsection{Probability of association: global computation}
\label{global}
We want to compute%
\footnote{%
  For the sake of clarity, we mention that we adopt the same 
  decreasing order of precedence for operators as in \textsc{Mathematica} 
  \citep{Mathematica}:
  $\times$ and $/$; $\prod$; $\sum$; $+$ and $-$.
}
the probability $\Prob(A_{i\comma j} \mid C \cap C')$ of association between 
sources $M_i$ and $\Mp_j$ ($j \neq 0$) or the probability that $M_i$ has 
no counterpart ($j = 0$), knowing the coordinates of all the objects in $K$ 
and $K'$.
Remembering that, for any events $\omega_1$, $\omega_2$, and $\omega_3$,
$\Prob(\omega_1 \mid \omega_2) = \Prob(\omega_1 \cap \omega_2)/\Prob(\omega_2)$
and thus
\begin{equation}
  \Prob(\omega_1 \cap \omega_2 \mid \omega_3)
  = \frac{\Prob(\omega_1 \cap \omega_2 \cap \omega_3)}{\Prob(\omega_3)}
  = \frac{\Prob(\omega_1 \mid \omega_2 \cap \omega_3)\multspace
    \Prob(\omega_2 \cap \omega_3)}{\Prob(\omega_3)}
  = \Prob(\omega_1 \mid \omega_2 \cap \omega_3) \multspace
  \Prob(\omega_2 \mid \omega_3),
  \label{Bayes2}
\end{equation}
we have,
with $\omega_1 = A_{i\comma j}$, $\omega_2 = C$, and $\omega_3 = C'$,
\begin{equation}
  \label{P(Aij|C,C')_gen}
  \Prob(A_{i\comma j} \mid C \cap C')
  = \frac{
    \Prob(A_{i\comma j} \cap C \mid C')
  }{
    \Prob(C \mid C')
  }.
\end{equation}
\subsubsection{Computation of \texorpdfstring{$\Psto(C \mid C')$}{\$P\_\{s:o\}(C | C')\$}}
%
We first compute the denominator of Eq.~\eqref{P(Aij|C,C')_gen}%
\footnote{%
  Computing $\Psto(C \mid C')$ is easier than for $\Psto(C' \mid C)$:
  the latter would require calculating 
  $\Psto(c'_{\smash[t]{\ell}} \mid \bigcap_{\smash[t]{k=1{;}\, j_k=\ell}}^n
  {[c_k \cap A_{k\comma j_k}]})$
  (cf.~Eq.~\eqref{jl_non_nul}) because several $M_k$ might be associated 
  with the same $\Mp_\ell$.
  This does not matter for computations made under assumption~\ref{H_oto}.
}.
The event
\begin{equation}
  \bigcap_{k=1}^n\bigcup_{j_k=0}^\np A_{k\comma j_k}
  =
  \bigcup_{j_1=0}^\np\bigcup_{j_2=0}^\np\cdots\bigcup_{j_n=0}^\np
  \bigcap_{k=1}^n A_{k\comma j_k}
\end{equation}
is certain by definition of the $A_{k\comma j_k}$ and, under either 
assumption~\ref{H_sto} or \ref{H_oto}, 
$A_{k\comma\ell} \cap A_{k\comma m} = \varnothing$ for all $M_k$ if $\ell \neq m$.
Consequently, using the symbol $\biguplus$ for mutually exclusive events
instead of $\bigcup$, we obtain
\begin{align}
  \Psto(C \mid C')
  &=
  \Psto\Bigl(C \cap \bigcap_{k=1}^n\bigcup_{j_k=0}^\np A_{k\comma j_k}
  \mid C'\Bigr)
  =
  \Psto\Bigl(C \cap
  \biguplus_{j_1=0}^\np\biguplus_{j_2=0}^\np\cdots\biguplus_{j_n=0}^\np
  \bigcap_{k=1}^n A_{k\comma j_k} \Bigm| C' \Bigr)
  = 
  \sum_{j_1=0}^\np \sum_{j_2=0}^\np \cdots \sum_{j_n=0}^\np
  \Psto\Bigl(C \cap \bigcap_{k=1}^n A_{k\comma j_k}
  \Bigm| C' \Bigr)
  \notag \\
  &= 
  \sum_{j_1=0}^\np \sum_{j_2=0}^\np \cdots \sum_{j_n=0}^\np
  \Psto\Bigl(C \Bigm| \bigcap_{k=1}^n A_{k\comma j_k}
  \cap C' \Bigr) \multspace
  \Psto\Bigl(\bigcap_{k=1}^n A_{k\comma j_k} \Bigm| C' \Bigr),
  \label{P_sto(C|C')_gen}
\end{align}
with $\omega_1 = C$, $\omega_2 = \bigcap_{\smash[t]{k=1}}^n A_{k\comma j_k}$,
and $\omega_3 = C'$ in Eq.~\eqref{Bayes2}.

Since $C = \bigcap_{\smash[t]{k=1}}^n c_k$,
the first factor in the product of Eq.~\eqref{P_sto(C|C')_gen} is
\begin{equation}
  \Pato\Bigl(C \Bigm| \bigcap_{k=1}^n A_{k\comma j_k} \cap C'\Bigr)
  =
  \Pato\Bigl(c_1 \Bigm| \bigcap_{k=2}^n c_k
  \cap \bigcap_{k=1}^n A_{k\comma j_k} \cap C'\Bigr)
  \multspace \Pato\Bigl(\bigcap_{k=2}^n c_k
  \Bigm| \bigcap_{k=1}^n A_{k\comma j_k} \cap C' \Bigr),
\end{equation}
with $\omega_1 = c_1$, $\omega_2 = \bigcap_{\smash[t]{k=2}}^n c_k$,
and $\omega_3 = A_{k\comma j_k} \cap C'$ in Eq.~\eqref{Bayes2}.
Doing the same with $\bigcap_{\smash[t]{k=2}}^n c_k$ instead of $C$, 
we obtain
\begin{equation}
  \Pato\Bigl(C \Bigm| \bigcap_{k=1}^n A_{k\comma j_k} \cap C'\Bigr)
  = 
  \prod_{\ell=1}^n \Pato\Bigl(c_\ell \Bigm| \bigcap_{k=\ell+1}^n c_k
  \cap \bigcap_{k=1}^n A_{k\comma j_k} \cap C'\Bigr)
  \label{P_sto(C|A,C')_gen}
\end{equation}
by iteration.

If $j_\ell \neq 0$, $M_\ell$ is only associated with $\Mp_{j_\ell}$.
Consequently,
\begin{equation}
  \Pato\Bigl(c_\ell \Bigm| \bigcap_{k=\ell+1}^n c_k
  \cap \bigcap_{k=1}^n A_{k\comma j_k} \cap C'\Bigr)
  = 
  \Pato(c_\ell \mid A_{\ell\comma j_\ell} \cap c'_{\smash[t]{j_\ell}})
  =
  \xi_{\ell\comma j_\ell} \multspace \df^2\vec r_\ell,
  \label{jl_non_nul}
\end{equation}
where,
denoting by $\vec r_{\ell\comma j_\ell} \coloneqq \vec r'_{\smash[t]{j_\ell}} - \vec r_\ell$
the position vector of $\Mp_{j_\ell}$ relative to $M_\ell$ and by
$\Gamma_{\ell\comma j_\ell}$ the covariance matrix of $\vec r_{\ell\comma j_\ell}$
(cf.~App.~\ref{cov_mat}),
\begin{equation}
  \xi_{\ell\comma j_\ell} 
  =
  \frac{
    \exp\Bigl(
    -\frac{1}{2}\multspace \transpose{\vec r}_{\smash[t]{\ell\comma j_\ell}} \cdot
    \Gamma_{\smash[t]{\ell\comma j_\ell}}^{-1} \cdot \vec r_{\ell\comma j_\ell}
    \Bigr)
  }{
    2\multspace \piup\multspace \!\sqrt{\det \Gamma_{\ell\comma j_\ell}}
  }.
\end{equation}

If $j_\ell = 0$, $M_\ell$ is not associated with any source in $K'$.
Since clustering is neglected,
\begin{equation}
  \Pato\Bigl(c_\ell \Bigm| \bigcap_{k=\ell+1}^n c_k \cap
  \bigcap_{k=1}^\np c'_{\smash[t]{k}} \cap \bigcap_{k=1}^n A_{k\comma j_k}\Bigr)
  = 
  \Pato(c_\ell \mid A_{\ell\comma 0})
  =
  \xi_{\ell\comma 0}\multspace \df^2\vec r_\ell,
  \label{jl_nul}
\end{equation}
where the last equality defines the spatial probability density
$\xi_{\ell\comma 0}$;
for the uninformative prior of a uniform \emph{a priori} probability
distribution of $K$-sources without counterpart, $\xi_{\ell\comma 0} = 1/\Stot$.

From Eqs.~\eqref{P_sto(C|A,C')_gen}, \eqref{jl_non_nul}, and \eqref{jl_nul},
it follows that
\begin{equation}
  \label{prod_xi}
  \Pato\Bigl(C \Bigm|
  \bigcap_{k=1}^n A_{k\comma j_k} \cap C' \Bigr)
  = 
  \Xi \multspace \prod_{k=1}^n \xi_{k\comma j_k},
\end{equation}
where
\begin{equation}
  \label{lambda}
  \Xi \coloneqq \prod_{k=1}^n \df^2\vec r_k.
\end{equation}

We now compute the second factor in the product of Eq.~\eqref{P_sto(C|C')_gen}.
Knowing the coordinates of $K'$-sources alone, without those of any in $K$, 
does not change the likelihood of the associations $(A_{k\comma j_k})$;
in other words, $C'$ and $\bigcap_{\smash[t]{k=1}}^n A_{k\comma j_k}$ are mutually
unconditionally independent (but conditionally dependent on $C$).
Therefore,
\begin{equation}
  \Psto\Bigl(\bigcap_{k=1}^n A_{k\comma j_k} \Bigm| C'\Bigr) =
  \Psto\Bigl(\bigcap_{k=1}^n A_{k\comma j_k}\Bigr).
\end{equation}
Let $q \coloneqq \card \{k \in \integinterv{1}{n} \mid j_k \neq 0\}$,
where $\card E$ denotes the number of elements of any set $E$.
Since the events $(A_{k\comma j_k})_{k\in\integinterv{1}{n}}$ are independent
by assumption~\ref{H_sto},
\begin{equation}
  \Psto\Bigl(\bigcap_{k=1}^n A_{k\comma j_k}\Bigr) 
  =
  \prod_{k=1}^n \Psto(A_{k\comma j_k}).
\end{equation}
Using definition~\eqref{def_f}, and on the hypothesis that all associations
$(A_{k\comma\ell})_{\ell\in\integinterv{1}{\np}}$ are \emph{a priori}
equally likely if $k \neq 0$ (Sect.~\ref{assump}), we get
\begin{equation}
  \Psto(A_{k\comma j_k}) 
  =
  \frac{\Psto(\bigcup_{\ell\neq0} A_{k\comma\ell})}{\card K'} = \frac{f}\np
  \quad\text{for } j_k \neq 0.
\end{equation}
Since $\Psto(A_{k\comma 0}) = 1-f$, we have
\begin{equation}
  \label{P_sto(A)}
  \Psto\Bigl(\bigcap_{k=1}^n A_{k\comma j_k}\Bigr)
  = 
  \Biggl(\frac{f}\np\Biggr)^q\multspace (1-f)^{n-q}.
\end{equation}

Hence, from Eqs.~\eqref{P_sto(C|C')_gen}, \eqref{prod_xi} and \eqref{P_sto(A)},
\begin{equation}
  \Psto(C \mid C')
  = 
  \Xi\multspace \sum_{j_1=0}^\np \sum_{j_2=0}^\np \cdots \sum_{j_n=0}^\np
  {\Biggl(\frac{f}\np\Biggr)^q \multspace (1-f)^{n-q}
    \multspace \prod_{k=1}^n \xi_{k\comma j_k}}.
  \label{P_sto(C|C')_xi}
\end{equation}
By the definition of $q$, there are $q$ strictly positive indices $j_k$
(as many as the factors ``$f/\np$'' in Eq.~\eqref{P_sto(C|C')_xi}) and
$n-q$ null ones (as many as the factors ``$(1-f)$'').
Therefore, with
\begin{equation}
  \label{def_zeta}
  \zeta_{k\comma 0} \coloneqq (1-f)\multspace \xi_{k\comma 0}
  \qquad\text{and}\qquad
  \zeta_{k\comma j_k} \coloneqq \frac{f\multspace \xi_{k\comma j_k}}{\np}
  \quad\text{for }j_k \neq 0,
\end{equation}
Eq.~\eqref{P_sto(C|C')_xi} reduces to
\begin{equation}
  \Psto(C \mid C')
  = \Xi\multspace \sum_{j_1=0}^\np\sum_{j_2=0}^\np \cdots
  \sum_{j_n=0}^\np\prod_{k=1}^n\zeta_{k\comma j_k}
  = \Xi\multspace \prod_{k=1}^n\sum_{j_k=0}^\np\zeta_{k\comma j_k},
  \label{P_sto(C|C')_res}
\end{equation}
where the last equality is derived by induction from the distributivity
of multiplication over addition.
\subsubsection{Computation of \texorpdfstring{$\Psto(A_{i\comma j} \cap C \mid C')$}{\$P\_\{s:o\}(A\_\{i, j\} \textbackslash cap C | C')\$}}
%
The computation of the numerator of Eq.~\eqref{P(Aij|C,C')_gen}
is similar to that of $\Psto(C \mid C')$:
\begin{align}
  \Psto(A_{i\comma j} \cap C \mid C')
  &=
  \Psto\Bigl(C \cap A_{i\comma j} \cap
  \biguplus_{j_1=0}^\np \cdots \biguplus_{j_{i-1}=0}^\np
  \biguplus_{j_{i+1}=0}^\np \cdots \biguplus_{j_n=0}^\np
  \bigcap_{\substack{k=1\\ k\neq i}}^n A_{k\comma j_k} \Bigm| C' \Bigr)
  = 
  \Psto\Bigl(C \cap \biguplus_{j_1=0}^\np \cdots \biguplus_{j_{i-1}=0}^\np
  \biguplus_{j_{i+1}=0}^\np \cdots \biguplus_{j_n=0}^\np
  \bigcap_{k=1}^n A_{k\comma j_k} \Bigm| C' \Bigr)
  \notag \\
  &=
  \sum_{j_1=0}^\np \cdots \sum_{j_{i-1}=0}^\np
  \sum_{j_{i+1}=0}^\np \cdots \sum_{j_n=0}^\np
  \Psto\Bigl(C \Bigm| \bigcap_{k=1}^n A_{k\comma j_k}
  \cap C' \Bigr) \multspace
  \Psto\Bigl(\bigcap_{k=1}^n A_{k\comma j_k} \Bigm| C' \Bigr),
  \label{P_sto(Aij,C|C')_gen}
\end{align}
where we put $j_i \coloneqq j$.

Let $q^\star \coloneqq \card\{k \in \integinterv{1}{n} \mid j_k \neq 0\}$
(indices $j_k$ are now those of Eq.~\eqref{P_sto(Aij,C|C')_gen}).
As for $\Psto(C \mid C')$,
\begin{align}
  \Psto(A_{i\comma j} \cap C \mid C')
  &=
  \Xi\multspace \sum_{j_1=0}^\np \cdots
  \sum_{j_{i-1}=0}^\np \sum_{j_{i+1}=0}^\np
  \cdots \sum_{j_n=0}^\np
  {\Biggl(\frac{f}\np\Biggr)^{q^\star}\multspace (1-f)^{n-q^\star}
    \prod_{k=1}^n \xi_{k\comma j_k}}
  = 
  \Xi\multspace \zeta_{i\comma j_i}\multspace \sum_{j_1=0}^\np\cdots
  \sum_{j_{i-1}=0}^\np\sum_{j_{i+1}=0}^\np\cdots
  \sum_{j_n=0}^\np\prod_{\substack{k=1\\ k\neq i}}^n \zeta_{k\comma j_k}
  \notag \\
  &=
  \Xi\multspace \zeta_{i\comma j}\multspace \prod_{\substack{k=1\\ k\neq i}}^n
  \sum_{j_k=0}^\np\zeta_{k\comma j_k}.
  \label{P_sto(Aij,C|C')_res}
\end{align}
\subsubsection{Final results}
Finally, from Eqs.~\eqref{P(Aij|C,C')_gen}, \eqref{P_sto(C|C')_res},
and \eqref{P_sto(Aij,C|C')_res},
\begin{align}
  \label{P_sto(Aij|C,C')_res1}
  \text{for } i \neq 0, \quad
  \Psto(A_{i\comma j} \mid C \cap C')
  &=
  \frac{
    \zeta_{i\comma j}\multspace \prod_{\leftsubstack{k=1\\ k\neq i}}^n
    \sum_{j_k=0}^\np\zeta_{k\comma j_k}
  }{
    \prod_{k=1}^n\sum_{j_k=0}^\np\zeta_{k\comma j_k}
  }
  = \frac{\zeta_{i\comma j}}{\sum_{k=0}^\np\zeta_{i\comma k}}
  \\
  \label{P_sto(Aij|C,C')_res2}
  &=
  \Left\{
  \begin{aligned}
    \frac{f\multspace \xi_{i\comma j}}{
      (1-f)\multspace \np\multspace \xi_{i\comma 0} 
      + f\multspace \sum_{k=1}^\np\xi_{i\comma k}}
    &
    \quad \text{for } j \neq 0,
    \\
    \frac{(1-f)\multspace \np\multspace \xi_{i\comma 0}}{
      (1-f)\multspace \np\multspace \xi_{i\comma 0} 
      + f\multspace \sum_{k=1}^\np\xi_{i\comma k}}
    &
    \quad \text{for } j = 0.
  \end{aligned}
  \Right.
\end{align}

As to the probability $\Psto(A_{0\comma j} \mid C \cap C')$ that $\Mp_j$
has no counterpart in $K$, it can be computed in this way:
\begin{align}
  \Psto(A_{0\comma j} \cap C \mid C')
  &=
  \Psto\Bigl(C \cap A_{0\comma j} \cap
  \biguplus_{j_1=0}^\np \biguplus_{j_2=0}^\np \cdots \biguplus_{j_n=0}^\np
  \bigcap_{k=1}^n A_{k\comma j_k} \Bigm| C'\Bigr)
  =
  \Psto\Bigl(C \cap \biguplus_{\substack{j_1=0\\ j_1\neq j}}^\np
  \biguplus_{\substack{j_2=0\\ j_2\neq j}}^\np
  \cdots \biguplus_{\substack{j_n=0\\ j_n\neq j}}^\np
  \bigcap_{k=1}^n A_{k\comma j_k} \Bigm| C'\Bigr)
  \notag \\
  &=
  \sum_{\substack{j_1=0\\ j_1\neq j}}^\np \sum_{\substack{j_2=0\\ j_2\neq j}}^\np
  \cdots \sum_{\substack{j_n=0\\ j_n\neq j}}^\np
  \Psto\Bigl(C \cap \bigcap_{k=1}^n A_{k\comma j_k}
  \Bigm| C'\Bigr)
  =
  \Xi\multspace \sum_{\substack{j_1=0\\ j_1\neq j}}^\np
  \sum_{\substack{j_2=0\\ j_2\neq j}}^\np
  \cdots \sum_{\substack{j_n=0\\ j_n\neq j}}^\np
  \prod_{k=1}^n\zeta_{k\comma j_k}
  =
  \Xi\multspace \prod_{k=1}^n\sum_{\substack{j_k=0\\ j_k\neq j}}^\np\zeta_{k\comma j_k}
\end{align}
and, using Eqs.~\eqref{P_sto(C|C')_res}, \eqref{P_sto(Aij|C,C')_res1},
and \eqref{Bayes2},
\begin{align}
  \Psto(A_{0\comma j} \mid C \cap C')
  &=
  \frac{
    \Psto(A_{0\comma j} \cap C \mid C')
  }{
    \Psto(C \mid C')
  }
  =
  \frac{\Xi\multspace \prod_{k=1}^n\sum_{\leftsubstack{j_k=0\\ j_k\neq j}}^\np
    \zeta_{k\comma j_k}}{\Xi\multspace \prod_{k=1}^n\sum_{j_k=0}^\np\zeta_{k\comma j_k}}
  =
  \prod_{k=1}^n\frac{\sum_{j_k=0}^\np\zeta_{k\comma j_k} - \zeta_{k\comma j}}{
    \sum_{j_k=0}^\np\zeta_{k\comma j_k}}
  =
  \prod_{k=1}^n{\Biggl(
    1-\frac{\zeta_{k\comma j_k}}{\sum_{j_k=0}^\np\zeta_{k\comma j_k}}
    \Biggr)}
  \notag \\
  &=
  \prod_{k=1}^n{\Bigl(1 - \Psto[A_{k\comma j} \mid C \cap C']\Bigr)}
  \quad \text{for } j \neq 0.
  \label{P_sto(A0j|C,C')}
\end{align}
\subsection{Likelihood and estimation of unknown parameters}
\label{fractionsto}
\subsubsection{General results}
Various methods have been proposed for estimating the fraction of
sources with a counterpart \citep{Kim, Fleuren, McAlpine, Haakonsen}.
\citet{Pineau}, for instance, fit $f$ to the overall distribution
of the likelihood ratios.
We propose a more convenient and systematic method in this section.

Besides $f$, the probabilities $\Prob(A_{i\comma j} \mid C \cap C')$ may depend 
on other unknowns, such as the parameters $\sigmatot$ and $\nutot$ modeling 
the positional uncertainties (cf.~Apps.~\ref{unknown} and \ref{different}).
We write here $x_1$, $x_2$, etc., for all these parameters, and put
$\vec x \coloneqq (x_1, x_2, \ldots)$.
An estimate $\hat{\vec x}$ of $\vec x$ may be obtained by maximizing
with respect to $\vec x$ (and with the constraint $\hat f \in[0, 1]$)
the overall likelihood
\begin{equation}
  \label{def_Lh}
  \Lh \coloneqq \frac{\Prob(C \cap C')}{
    (\prod_{i=1}^n \df^2\vec r_i)\multspace \prod_{j=1}^\np\df^2\vec r'_{\smash[t]{j}}}
\end{equation}
to observe all the $K$-~and $K'$-sources at their effective positions.
Unless the result is outside the possible domain for $\vec x$
(i.e., if $\Lh$ reaches its maximum on the boundary of this domain),
the maximum likelihood estimator $\hat{\vec x}$ is a solution to
\begin{equation}
  \label{max_Lh}
  \Biggl(\frac{\partial\ln\Lh}{\partial\vec x}\Biggr)_{\vec x=\hat{\vec x}} = 0.
\end{equation}
From now on,
all quantities calculated at $\vec x = \hat{\vec x}$ bear a circumflex.

We have
\begin{equation}
  \label{P(C,C')}
  \Prob(C \cap C') = \Prob(C \mid C')\multspace \Prob(C'),
\end{equation}
and, since clustering is neglected,
\begin{equation}
  \label{P(C')}
  \Prob(C') = \prod_{j=1}^\np \Prob(\coordpj)
  = 
  \prod_{j=1}^\np \xi_{0\comma j}\multspace \df^2\vec r'_{\smash[t]{j}},
\end{equation}
where $\xi_{0\comma j}$ is the spatial probability density defined by
$\Prob(\coordpj) = \xi_{0\comma j}\multspace \df^2\vec r'_{\smash[t]{j}}$;
for the uninformative prior of a uniform \emph{a priori}
probability distribution of $K'$-sources, $\xi_{0\comma j} = 1/\Stot$.
From Eqs.~\eqref{def_Lh}, \eqref{P(C,C')}, \eqref{P(C')}, and \eqref{lambda},
we obtain
\begin{equation}
  \label{Lh_gen}
  \Lh = \frac{\Prob(C \mid C')}{\Xi}\multspace
  \prod_{j=1}^\np \xi_{0\comma j}.
\end{equation}

In particular, under assumption~\ref{H_sto}, Eqs.~\eqref{Lh_gen}, 
\eqref{P_sto(C|C')_res}, and \eqref{lambda} give
\begin{equation}
  \label{Lh_sto}
  \Lhsto = \Bigl(\prod_{i=1}^n\sum_{k=0}^\np\zeta_{i\comma k}\Bigr)\multspace
  \prod_{j=1}^\np \xi_{0\comma j}.
\end{equation}
Therefore, for any parameter $x_p$ and because the $\xi_{0\comma j}$ 
are independent of $\vec x$,
\begin{equation}
  \frac{\partial\ln\Lhsto}{\partial x_p}
  =
  \sum_{i=1}^n\frac{\partial\ln\sum_{k=0}^\np\zeta_{i\comma k}}{\partial x_p}
  = 
  \sum_{i=1}^n\sum_{j=0}^\np\frac{\partial\zeta_{i\comma j}/\partial x_p}{
    \sum_{k=0}^\np\zeta_{i\comma k}}
  = 
  \sum_{i=1}^n\sum_{j=0}^\np\frac{\partial\ln\zeta_{i\comma j}}{\partial x_p}
  \multspace \frac{\zeta_{i\comma j}}{\sum_{k=0}^\np\zeta_{i\comma k}}
  =
  \sum_{i=1}^n\sum_{j=0}^\np\frac{\partial\ln\zeta_{i\comma j}}{\partial x_p}
  \multspace \Psto(A_{i\comma j} \mid C \cap C').
  \label{der(Lh_sto)/x}
\end{equation}
(For reasons highlighted just after Eq.~\eqref{Lh_oto_brut}, it is convenient 
to express most results as a function of the probabilities
$\Prob(A_{i\comma j} \mid C \cap C')$.)

Uncertainties on the unknown parameters may be computed from the covariance
matrix $V$ of $\hat{\vec x}$.
For large numbers of sources, $V$ is asymptotically given \citep{KS} by
\begin{equation}
  \label{cov_x}
  \Bigl(V^{-1}\Bigr)_{p\comma q}
  = -\Left(
  \frac{\partial^2\ln\Lh}{\partial x_p\multspace \partial x_q}
  \Right)_{\hat{\vec x} = \vec x}.
\end{equation}
\subsubsection{Fraction of sources with a counterpart}
Consider, in particular, the case $x_p = f$.
We note that
\begin{equation}
  \label{der(zeta)}
  \frac{\partial\ln\zeta_{i\comma 0}}{\partial f} = -\frac{1}{1-f}
  \qquad\text{and}\qquad
  \frac{\partial\ln\zeta_{i\comma j}}{\partial f} = \frac{1}{f}
  \quad\text{for } j \neq 0.
\end{equation}
Under the assumption~\ref{H_sto} or \ref{H_oto} (but not under
\ref{H_ots}),
\begin{equation}
  \label{somme_prob}
  \sum_{j=0}^\np \Pato(A_{i\comma j} \mid C \cap C') 
  =
  \Pato\Bigl(\biguplus_{j=0}^\np A_{i\comma j} \Bigm| C \cap C'\Bigr) = 1,
\end{equation}
so, using Eq.~\eqref{der(zeta)},
\begin{align}
  \sum_{j=0}^\np\frac{\partial\ln\zeta_{i\comma j}}{\partial f}\multspace
  \Pato(A_{i\comma j} \mid C \cap C')
  &=
  -\frac{\Pato(A_{i\comma 0} \mid C \cap C')}{1-f}
  + \sum_{j=1}^\np \frac{\Pato(A_{i\comma j} \mid C \cap C')}{f}
  =
  -\frac{\Pato(A_{i\comma 0} \mid C \cap C')}{1-f}
  + \frac{1-\Pato(A_{i\comma 0} \mid C \cap C')}{f}
  \notag \\
  &=
  \frac{(1-f) - \Pato(A_{i\comma 0} \mid C \cap C') }{
    f\multspace (1-f)}.
  \label{somme_j}
\end{align}
Summing Eq.~\eqref{somme_j} on $i$, we obtain
from Eq.~\eqref{der(Lh_sto)/x} that
\begin{equation}
  \label{der(Lh_sto)/f}
  \frac{\partial\ln\Lhsto}{\partial f}
  = \frac{
    n\multspace (1-f) - \sum_{i=1}^n\Psto(A_{i\comma 0} \mid C \cap C')
  }{
    f\multspace (1-f)
  }.
\end{equation}

Consequently, the maximum likelihood estimator
of the fraction $f$ of $K$-sources with a counterpart in $K'$ is
\begin{align}
  \hat f_\sto
  &=
  1 - \frac{1}{n}\multspace
  \sum_{i=1}^n\expandafter\hat\Psto(A_{i\comma 0} \mid C \cap C')
  \label{f_est_sto1}
  \\
  &=
  \frac{1}{n}\multspace
  \sum_{i=1}^n\sum_{j=1}^\np\expandafter\hat\Psto(A_{i\comma j}
  \mid C \cap C').
  \label{f_est_sto2}
\end{align}
After some tedious calculations, it can be shown that
\begin{equation}
  \label{concave}
  \frac{\partial^2\ln\Lhsto}{\partial f^2}
  = -\frac{
    \sum_{i=1}^n{\Bigl([1-f] -
      \Psto[A_{i\comma 0} \mid C \cap C']\Bigr)^2}
  }{
    f^2\multspace (1-f)^2
  }
  < 0
\end{equation}
for all $f$, so $\partial\ln\Lhsto/\partial f$ has at most one zero in $[0, 1]$:
$\hat f_\sto$ is unique.

Since $\hat f_\sto$ appears on the two sides of Eq.~\eqref{f_est_sto1} (remember 
that $\expandafter\hat\Psto$ is the value of $\Psto$ at $f = \hat f_\sto$),
we may try to determine it through an iterative back and forth computation
between the lefthand and the righthand sides of this equation.
(A similar idea was also proposed by \citealt{Benn}.)
We prove in Sect.~\ref{cvg_f} that this procedure converges
for any starting value $f \in \mathopen]0, 1\mathclose[$.

An estimate $\hat f'_\sto$ of the fraction $f'$ of $K'$-sources with a 
counterpart is given by
\begin{equation}
  \hat f'_\sto 
  = 
  1 - \frac{1}{\np}\multspace
  \sum_{j=1}^\np\expandafter\hat\Psto(A_{0\comma j} \mid C \cap C').
  \label{f'_est_sto}
\end{equation}
It can be checked from Eqs.~\eqref{f_est_sto2}, \eqref{f'_est_sto}, 
and \eqref{P_sto(A0j|C,C')} that, as expected if assumption~\ref{H_sto} is 
valid (cf.~Sect~\ref{assump}),
$\hat f_\sto\multspace n \geqslant \hat f'_\sto\multspace \np$.
(Just notice that, for any numbers $y_i \in [0, 1]$,
$\prod_{i=1}^n {(1 - y_i)} \geqslant 1 - \sum_{i=1}^n y_i$,
which is obvious by induction;
apply this to $y_i = \expandafter\hat\Psto(A_{i\comma j} \mid C \cap C')$
and then sum on $j$.)
\subsection{Probability of association: local computation}
\label{local}
Under assumption~\ref{H_sto}, a purely local computation (subscript ``\loc''
hereafter) of the probabilities of association is also possible.
Consider a region $U_i$ of area $\Si$ containing the position 
of $M_i$, and such that 
we
can safely hypothesize that the counterpart 
in $K'$ of $M_i$, if any, is inside.
We assume that the local surface density $\rhopi$ of $K'$-sources unrelated 
to $M_i$ is uniform on $U_i$.
To avoid biasing the estimate if $M_i$ has a counterpart, $\rhopi$ may be 
evaluated from the number of $K'$-sources in a region surrounding $U_i$, but 
not overlapping it (an annulus around a disk $U_i$ centered on $M_i$, 
for instance).

Besides the $A_{i\comma j}$, we consider the following events:
\begin{itemize}
\item
  $\Npi$: $U_i$ contains $\npi$ sources;
\item
  $\COORDpi \coloneqq \bigcap_{j \in J_i} \coordpj$,
  where $J_i \coloneqq \{j \mid \Mp_j \in U_i\}$.
\end{itemize}

We want to compute the probability that a source $\Mp_j$ in $U_i$ is the
counterpart of $M_i$, given the positions relative to $M_i$ of
all its possible counterparts $(\Mp_k)^{}_{\smash[t]{k\in J_i}}$,
i.e. $\Ploc(A_{i\comma j} \mid \COORDpi\cap \Npi)$.
Using Eq.~\eqref{Bayes2} with $\omega_1 = A_{i\comma j}$,
$\omega_2 = \COORDpi$, and $\omega_3 = \Npi$ in the first equality below,
and then with $\omega_1 = \COORDpi$, $\omega_2 = A_{i\comma k}$, and
$\omega_3$ unchanged in the last one, 
we obtain
\begin{align}
  \Ploc(A_{i\comma j} \mid \COORDpi\cap \Npi)
  &
  = 
  \frac{\Ploc(A_{i\comma j} \cap \COORDpi \mid \Npi)}{
    \Ploc(\COORDpi \mid \Npi)}
  = 
  \frac{\Ploc(\COORDpi \cap A_{i\comma j} \mid \Npi)}{
    \Ploc(\COORDpi \cap \biguplus_{k\in J_i\cup\{0\}} A_{i\comma k} \mid \Npi)}
  =
  \frac{\Ploc(\COORDpi \cap A_{i\comma j} \mid \Npi)}{
    \sum_{k\in J_i\cup\{0\}} \Ploc(\COORDpi \cap A_{i\comma k} \mid \Npi)}
  \notag \\
  &
  =
  \frac{\Ploc(\COORDpi\mid A_{i\comma j} \cap \Npi)\multspace
    \Ploc(A_{i\comma j} \mid \Npi)}{
    \sum_{k\in J_i\cup\{0\}} \Ploc(\COORDpi\mid A_{i\comma k} \cap \Npi)
    \multspace \Ploc(A_{i\comma k} \mid \Npi)}.
  \label{P_loc(Aij|C',N')}
\end{align}

Now,
\begin{equation}
  \label{P_loc(Ai0|N')}
  \Ploc(A_{i\comma 0} \mid \Npi) 
  =
  \frac{\Ploc( \Npi \cap A_{i\comma 0})}{\Ploc(\Npi)} 
  =
  \frac{\Ploc(\Npi\mid A_{i\comma 0})\multspace
    \Ploc(A_{i\comma 0})}{
    \Ploc(\Npi\mid A_{i\comma 0})\multspace \Ploc(A_{i\comma 0}) +
    \Ploc(\Npi\mid \overline{A_{i\comma 0}})\multspace
    \Ploc(\overline{A_{i\comma 0}})}
\end{equation}
and
\begin{equation}
  \label{P_loc(Aij|N')}
  \Ploc(A_{i\comma j} \mid \Npi)
  =
  \frac{\Ploc(\overline{A_{i\comma 0}} \mid \Npi)}{\npi}
  =
  \frac{1-\Ploc(A_{i\comma 0} \mid \Npi)}{\npi}
  \quad \text{for } j \neq 0.
\end{equation}
(The probability $\Ploc(A_{i\comma j})$ itself could not have been computed 
as $\Ploc(\overline{A_{i\comma 0}})/\npi$ because $\npi$ would be undefined,
which is why event $\Npi$ was introduced.)
If clustering is negligible, the number of $K'$-sources randomly distributed
with a mean surface density $\rhopi$ in an area $\Si$ follows a Poissonian 
distribution, so
\begin{equation}
  \label{P_loc(N'|nonAi0)}
  \Ploc(\Npi \mid \overline{A_{i\comma 0}})
  =
  \frac{(\rhopi\multspace \Si)^{\npi-1}\multspace \exp(-\rhopi\multspace \Si)}{
    (\npi-1)!}
  \qquad\text{(one counterpart and $\npi-1$ sources by chance in $\Si$)}
\end{equation}
and
\begin{equation}
  \label{P_loc(N'|Ai0)}
  \Ploc(\Npi\mid A_{i\comma 0})
  =
  \frac{(\rhopi\multspace \Si)^{\npi}\multspace \exp(-\rhopi\multspace \Si)}{
    \npi!}
  \qquad\text{(no counterpart and $\npi$ sources by chance in $\Si$).}
\end{equation}
Thus, from Eqs.~\eqref{P_loc(Aij|N')}, \eqref{P_loc(Ai0|N')}, 
\eqref{P_loc(N'|nonAi0)}, \eqref{P_loc(N'|Ai0)}, and \eqref{def_f},
\begin{equation}
  \label{P_loc(Aij|N')_res}
  \Ploc(A_{i\comma j} \mid \Npi)
  =
  \Left\{
  \begin{aligned}
    \frac{f}{\npi\multspace f+(1-f)\multspace \rhopi\multspace \Si}
    &
    \quad \text{for } j \neq 0,
    \\
    \frac{(1-f)\multspace \rhopi\multspace \Si}{
      \npi\multspace f+(1-f)\multspace \rhopi\multspace \Si}
    &
    \quad \text{for } j = 0.
  \end{aligned}
  \Right.
\end{equation}

We have
\begin{equation}
  \label{P_loc(C'|Aij,N')}
  \Ploc(\COORDpi\mid A_{i\comma 0}\cap \Npi) 
  =
  \prod_{k\in J_i} \frac{\df^2\vec r'_{\smash[t]{k}}}{\Si}
  \qquad\text{and}\qquad
  \Ploc(\COORDpi\mid A_{i\comma j} \cap \Npi) 
  =
  \xi_{i\comma j}\multspace \df^2\vrpj\multspace
  \prod_{\substack{k\in J_i\\ k\neq j}} \frac{\df^2\vec r'_{\smash[t]{k}}}{\Si}
  \quad\text{for } j \neq 0
\end{equation}
(rigorously, $\xi_{i\comma j}$ should be replaced by
$\xi_{i\comma j}/\Ploc(\Mp_j\in U_i\mid A_{i\comma j})$, but
$\Ploc(\Mp_j\not\in U_i \mid A_{i\comma j})$ is negligible by definition of $U_i$),
so, using Eqs.~\eqref{P_loc(Aij|C',N')}, \eqref{P_loc(Aij|N')_res}, and
\eqref{P_loc(C'|Aij,N')}, 
we obtain
\begin{equation}
  \label{P_loc}
  \Ploc(A_{i\comma j}\mid \COORDpi\cap \Npi)
  =
  \Left\{
  \begin{aligned}
    \frac{f\multspace \lambda_{i\comma j}}{
      (1-f)+f\multspace \sum_{k\in J_i}\lambda_{i\comma k}}
    &
    \quad \text{for } j \neq 0,
    \\
    \frac{(1-f)}{(1-f)+f\multspace \sum_{k\in J_i}\lambda_{i\comma k}}
    &
    \quad \text{for } j = 0,
  \end{aligned}
  \Right.
\end{equation}
where $\lambda_{i\comma k} \coloneqq \xi_{i\comma k}/\rhopi$ is the likelihood ratio
(cf.~Eq.~\eqref{def_LR}).
\emph{Mutatis mutandis}, 
we obtain the same result as Eq.~(14) of
\citet{Pineau} and the aforementioned authors.
When the computation is extended from $U_i$ to the 
\vadjust{\vspace{-1pt}}whole surface covered by $K'$,
$\rhopi$ is replaced by $\np\!/\Stot$ in Eq.~\eqref{P_loc}, $\sum_{k\in J_i}$ 
by $\sum_{k=1}^\np$, and we recover Eq.~\eqref{P_sto(Aij|C,C')_res2} 
since $\xi_{i\comma 0} = 1/\Stot$ for a uniform distribution.

The index $\jMLC(i)$ of the most likely counterpart $\Mp_{\jMLC(i)}$ of $M_i$ 
is the value of $j \neq 0$ maximizing $\lambda_{i\comma j}$.
Very often,
$\lambda_{i\comma \jMLC(i)} \gg \sum_{k\in J_i{;}\, k\neq \jMLC(i)} \lambda_{i\comma k}$,
so
\begin{equation}
  \Psto(A_{i\comma \jMLC(i)}\mid \COORDpi\cap \Npi)
  \approx 
  \frac{f\multspace \lambda_{i\comma \jMLC(i)}}{
    (1-f)+f\multspace \lambda_{i\comma \jMLC(i)}}.
\end{equation}
As a ``poor man's'' recipe, if the value of $f$ is unknown and not too
close to either $0$ or $1$, an association may be considered as true if 
$\lambda_{i\comma \jMLC(i)}\gg 1$ and as false if $\lambda_{i\comma \jMLC(i)}\ll 1$.
Where to set the boundary between true associations and false ones is
somewhat arbitrary \citep{Wolstencroft}.
For a large sample, however, $f$ can be estimated from the distribution
of the positions of all sources, as shown in Sect.~\ref{fractionsto}.
\section{One-to-one associations}
\label{one}
Under~\ref{H_sto} (Sect.~\ref{several}), a given $\Mp_j$ can be associated 
with several $M_i$:
there is no \vadjust{\vspace{-1pt}}symmetry between $K$ and $K'$ under this assumption and, while
$\sum_{\smash[t]{j=0}}^{\smash[t]{\np}} \Psto(A_{i\comma j}\mid C \cap C') = 1$
for all $M_i$, 
$\sum_{\smash[t]{i=1}}^n \Psto(A_{i\comma j}\mid C \cap C')$ could be strictly larger than $1$
for some sources $\Mp_j$.
We assume here that the much more constraining assumption~\ref{H_oto} holds.
As far as we know and despite some attempt by \citet{Rutledge},
this problem has not been solved previously
(see also \citealt{Bartlett} for a simple statement of the question).

Since a $K'$-potential counterpart $\Mp_j$ of $M_i$ within some neighborhood 
$U_i$ of $M_i$ might in fact be the true counterpart
of another source $M_k$ \emph{outside} of $U_i$,
there is no obvious way to adapt the exact local several-to-one
computation of Sect.~\ref{local} to the case of the one-to-one assumption.
We therefore have to consider 
all the $K$-~and $K'$-sources, as in Sect.~\ref{global}.

Under assumption~\ref{H_oto}, catalogs $K$ and $K'$ play symmetrical roles;
in particular,
\begin{equation}
  \label{f_f'_oto}
  \Poto(A_{i\comma j}) = \frac{f}\np = \frac{f'}n
  \quad \text{if $i \neq 0$ and $j \neq 0$}.
\end{equation}
For practical reasons (cf.~Eq.~\eqref{binom}), we nonetheless 
name $K$ the catalog with the fewer objects and $K'$ the other one,
so $n \leqslant \np$ in the following.
\subsection{Probability of association}
\label{sect_P_oto}
\subsubsection{Computation of \texorpdfstring{$\Poto(C \mid C')$}{\$P\_\{o:o\}(C | C')\$}}
%
The denominator of Eq.~\eqref{P(Aij|C,C')_gen} is
\begin{equation}
  \Poto(C \mid C')
  =  
  \Poto\Bigl(C \cap
  \biguplus_{j_1=0}^\np\biguplus_{j_2=0}^\np\cdots\biguplus_{j_n=0}^\np
  \bigcap_{k=1}^n A_{k\comma j_k} \Bigm| C' \Bigr)
\end{equation}
(same reasons as for Eq.~\eqref{P_sto(C|C')_gen}).
Because $A_{k\comma m} \cap A_{\ell\comma m} = \varnothing$ if
$k \neq \ell$ and $m \neq 0$ by assumption~\ref{H_oto}, this reduces to
\begin{equation}
  \label{P_oto(C|C')_union}
  \Poto(C \mid C')
  =  
  \Poto\Bigl(C \cap \biguplus_{\substack{j_1=0\\ j_1\not\in X_0}}^\np
  \biguplus_{\substack{j_2=0\\ j_2\not\in X_1}}^\np \cdots
  \biguplus_{\substack{j_n=0\\ j_n\not\in X_{n-1}}}^\np
  \bigcap_{k=1}^n A_{k\comma j_k} \Bigm| C' \Bigr),
\end{equation}
where, to ensure that each $K'$-source is associated with at most one of $K$,
the sets $X_k$ of excluded counterparts are defined iteratively by
\begin{equation}
  \label{def_J}
  X_0 \coloneqq \varnothing
  \qquad \text{and}\qquad
  X_k \coloneqq (X_{k-1} \cup \{j_k\}) \setminus \{0\}
  \quad\text{for all } k \in \integinterv{1}{n}.
\end{equation}
As a result,
\begin{equation}
  \Poto(C \mid C')
  =
  \sum_{\substack{j_1=0\\ j_1\not\in X_0}}^\np \sum_{\substack{j_2=0\\ j_2\not\in X_1}}^\np
  \cdots \sum_{\substack{j_n=0\\ j_n\not\in X_{n-1}}}^\np
  \Poto\Bigl(C \cap \bigcap_{k=1}^n A_{k\comma j_k} \Bigm| C' \Bigr)
  =
  \sum_{\substack{j_1=0\\ j_1\not\in X_0}}^\np \sum_{\substack{j_2=0\\ j_2\not\in X_1}}^\np
  \cdots \sum_{\substack{j_n=0\\ j_n\not\in X_{n-1}}}^\np
  \Poto\Bigl(C \Bigm| \bigcap_{k=1}^n A_{k\comma j_k} \cap C' \Bigr)
  \multspace \Poto\Bigl(\bigcap_{k=1}^n A_{k\comma j_k} \Bigm| C' \Bigr).
  \label{P_oto(C|C')_gen}
\end{equation}

The first factor in the product of Eq.~\eqref{P_oto(C|C')_gen} is still
given by Eq.~\eqref{prod_xi}, so 
we just have
to compute the second factor,
\begin{equation}
  \Poto\Bigl(\bigcap_{k=1}^n A_{k\comma j_k} \Bigm| C'\Bigr) 
  =
  \Poto\Bigl(\bigcap_{k=1}^n A_{k\comma j_k}\Bigr).
\end{equation}
Let $q \coloneqq \card X_n$ and $Q$ be a random variable describing 
the number of associations between $K$ and $K'$:
\begin{equation}
  \Poto\Bigl(\bigcap_{k=1}^n A_{k\comma j_k}\Bigr) 
  =
  \Poto\Bigl(\bigcap_{k=1}^n A_{k\comma j_k} \Bigm| Q = q\Bigr) \multspace
  \Poto(Q = q) +
  \Poto\Bigl(\bigcap_{k=1}^n A_{k\comma j_k} \Bigm| Q \neq q\Bigr) \multspace
  \Poto(Q \neq q).
\end{equation}
Since $\Poto(\bigcap_{k=1}^n A_{k\comma j_k} \mid Q \neq q) = 0$ by definition 
of $q$, 
we only have
to compute $\Poto(\bigcap_{k=1}^n A_{k\comma j_k} \mid Q = q)$ 
and $\Poto(Q = q)$.

There are $n!/(q!\multspace [n-q]!)$ choices of $q$ elements among $n$ in $K$,
and $\np!/(q!\multspace [\np-q]!)$ choices of $q$ elements among $\np$ in $K'$.
The number of permutations of $q$ elements is $q!$, so the total number of 
one-to-one associations of $q$ elements from $K$ to $q$ elements of $K'$ is
\begin{equation}
  q!\multspace \frac{n!}{q!\multspace (n-q)!}\multspace 
  \frac{\np!}{q!\multspace (\np-q)!}.
\end{equation}
The inverse of this number is
\begin{equation}
  \label{P_oto(A|m)}
  \Poto\Bigl(\bigcap_{k=1}^n A_{k\comma j_k} \Bigm| Q = q\Bigr) 
  =
  \frac{q!\multspace (n-q)!\multspace (\np-q)!}{n!\multspace \np!}.
\end{equation}

With our definition of $K$ and $K'$, $n \leqslant \np$, so all
the elements of $K$ may have a counterpart in $K'$ jointly.
Therefore, $\Poto(Q = q)$ is given by the binomial law:
\begin{equation}
  \label{binom}
  \Poto(Q = q) 
  = 
  \frac{n!}{q!\multspace (n-q)!}\multspace f^q\multspace (1-f)^{n-q}.
\end{equation}

From Eqs.~\eqref{P_oto(C|C')_gen}, \eqref{prod_xi}, \eqref{P_oto(A|m)}, and
\eqref{binom}, 
we obtain
\begin{align}
  \Poto(C \mid C')
  &=
  \Xi\multspace \sum_{\substack{j_1=0\\ j_1\not\in X_0}}^\np
  \sum_{\substack{j_2=0\\ j_2\not\in X_1}}^\np
  \cdots \sum_{\substack{j_n=0\\ j_n\not\in X_{n-1}}}^\np
  {\frac{(\np-q)!}{\np!}
    \multspace f^q \multspace (1-f)^{n-q}\multspace \prod_{k=1}^n \xi_{k\comma j_k}}
  \\
  &=
  \Xi\multspace \sum_{\substack{j_1=0\\ j_1\not\in X_0}}^\np
  \sum_{\substack{j_2=0\\ j_2\not\in X_1}}^\np
  \cdots \sum_{\substack{j_n=0\\ j_n\not\in X_{n-1}}}^\np
  {\Bigl(\prod_{\ell=1}^q\frac{f}{\np-\ell+1}\Bigr)\multspace
    \Bigl(\prod_{\ell=1}^{n-q}[1-f]\Bigr)\multspace
    \prod_{k=1}^n \xi_{k\comma j_k}}.
  \label{P_oto(C|C')_eta}
\end{align}
There are $q$ factors ``$f/(\np-\ell+1)$'' in the above equation,
one for each index $j_k \neq 0$.
There are also $n-q$ factors ``$(1-f)$'', one for each null $j_k$.
For every $j_k \neq 0$, 
$\card X_k = \card X_{k-1} + 1$;
and, since $q = \card X_n$, a different $j_k$ corresponds to each
$\ell \in \integinterv{1}{q}$, so $\ell = \card X_k$.
With
\begin{equation}
  \label{def_eta}
  \eta_{k\comma 0} \coloneqq \zeta_{k\comma 0}
  \qquad \text{and} \qquad
  \eta_{k\comma j_k} \coloneqq \frac{f\multspace \xi_{k\comma j_k}}{\np-\card X_{k-1}}
  \quad\text{for } j_k \neq 0,
\end{equation}
Eq.~\eqref{P_oto(C|C')_eta} therefore simplifies to
\begin{equation}
  \Poto(C \mid C')
  = 
  \Xi\multspace \sum_{\substack{j_1=0\\ j_1\not\in X_0}}^\np
  \sum_{\substack{j_2=0\\ j_2\not\in X_1}}^\np\cdots
  \sum_{\substack{j_n=0\\ j_n\not\in X_{n-1}}}^\np
  \prod_{k=1}^n \eta_{k\comma j_k}.
  \label{P_oto(C|C')_res}
\end{equation}
\subsubsection{Computation of \texorpdfstring{$\Poto(A_{i\comma j} \cap C \mid C')$}{\$P\_\{o:o\}(A\_\{i, j\} \textbackslash cap C | C')\$}}
%
The denominator of Eq.~\eqref{P(Aij|C,C')_gen} is computed in the same way
as $\Poto(C \mid C')$:
\begin{align}
  \Poto(A_{i\comma j} \cap C \mid C')
  &=
  \Poto\Bigl(C \cap A_{i\comma j} \cap
  \biguplus_{\substack{j_1=0\\ j_1\not\in X^\star_0}}^\np \cdots
  \biguplus_{\substack{j_{i-1}=0\\ j_{i-1}\not\in X^\star_{i-2}}}^\np
  \biguplus_{\substack{j_{i+1}=0\\ j_{i+1}\not\in X^\star_{i}}}^\np \cdots
  \biguplus_{\substack{j_n=0\\ j_n\not\in X^\star_{n-1}}}^\np
  \bigcap_{\substack{k=1\\ k\neq i}}^n A_{k\comma j_k} \Bigm| C' \Bigr)
  \notag \\
  &=
  \Poto\Bigl(C \cap \biguplus_{\substack{j_1=0\\ j_1\not\in X^\star_0}}^\np
  \cdots \biguplus_{\substack{j_{i-1}=0\\ j_{i-1}\not\in X^\star_{i-2}}}^\np
  \biguplus_{\substack{j_{i+1}=0\\ j_{i+1}\not\in X^\star_{i}}}^\np \cdots
  \biguplus_{\substack{j_n=0\\ j_n\not\in X^\star_{n-1}}}^\np
  \bigcap_{k=1}^n A_{k\comma j_k} \Bigm| C' \Bigr),
\end{align}
where
\begin{equation}
  \label{def_J*}
  X^\star_0 \coloneqq \{j\} \setminus \{0\},
  \qquad
  j_i \coloneqq j
  \qquad\text{and}\qquad
  X^\star_{k} \coloneqq (X^\star_{k-1} \cup \{j_k\}) \setminus \{0\}
  \quad\text{for all } k\in \integinterv{1}{n},
\end{equation}
so
\begin{equation}
  \Poto(A_{i\comma j} \cap C \mid C')
  = 
  \sum_{\substack{j_1=0\\ j_1\not\in X^\star_0}}^\np
  \cdots\sum_{\substack{j_{i-1}=0\\ j_{i-1}\not\in X^\star_{i-2}}}^\np
  \sum_{\substack{j_{i+1}=0\\ j_{i+1}\not\in X^\star_{i}}}^\np
  \cdots \sum_{\substack{j_n=0\\ j_n\not\in X^\star_{n-1}}}^\np
  \Poto\Bigl(C \Bigm| \bigcap_{k=1}^n A_{k\comma j_k}
  \cap C' \Bigr) \multspace
  \Poto\Bigl(\bigcap_{k=1}^n A_{k\comma j_k} \Bigm| C' \Bigr).
\end{equation}

Let $q^\star \coloneqq \card X^\star_n$.
As for $\Poto(C \mid C')$,
\begin{align}
  \Poto(A_{i\comma j} \cap C \mid C')
  &=
  \Xi\multspace \sum_{\substack{j_1=0\\ j_1\not\in X^\star_0}}^\np \cdots
  \sum_{\substack{j_{i-1}=0\\ j_{i-1}\not\in X^\star_{i-2}}}^\np
  \sum_{\substack{j_{i+1}=0\\ j_{i+1}\not\in X^\star_{i}}}^\np
  \cdots \sum_{\substack{j_n=0\\ j_n\not\in X^\star_{n-1}}}^\np
  {\frac{(\np-q^\star)!}{\np!} \multspace 
    f^{q^\star} \multspace (1-f)^{n-q^\star}\multspace \prod_{k=1}^n \xi_{k\comma j_k}}
  \notag \\
  &=
  \Xi\multspace \zeta_{i\comma j}\multspace
  \sum_{\substack{j_1=0\\ j_1\not\in X^\star_0}}^\np \cdots
  \sum_{\substack{j_{i-1}=0\\ j_{i-1}\not\in X^\star_{i-2}}}^\np
  \sum_{\substack{j_{i+1}=0\\ j_{i+1}\not\in X^\star_{i}}}^\np
  \cdots \sum_{\substack{j_n=0\\ j_n\not\in X^\star_{n-1}}}^\np
  \prod_{\substack{k=1\\ k\neq i}}^n \eta^\star_{k\comma j_k},
  \label{P_oto(Aij,C|C')_res}
\end{align}
where
\begin{equation}
  \label{def_eta*}
  \eta^\star_{k\comma 0} \coloneqq \zeta_{k\comma 0}
  \qquad\text{and}\qquad
  \eta^\star_{k\comma j_k} \coloneqq
  \frac{f\multspace \xi_{k\comma j_k}}{\np-\card X^\star_{k-1}}
  \quad \text{for } j_k \neq 0.
\end{equation}
\subsubsection{Final results}
Finally, from Eqs.~\eqref{P(Aij|C,C')_gen}, \eqref{P_oto(C|C')_res}, 
and \eqref{P_oto(Aij,C|C')_res},
\begin{equation}
  \label{P_oto(Aij|C,C')_res}
  \text{for } i \neq 0, \quad
  \Poto(A_{i\comma j} \mid C \cap C')
  = 
  \frac{
    \zeta_{i\comma j}\multspace
    \sum_{\leftsubstack{j_1=0\\ j_1\not\in X^\star_0}}^\np \cdots
    \sum_{\leftsubstack{j_{i-1}=0\\ j_{i-1}\not\in X^\star_{i-2}}}^\np
    \sum_{\leftsubstack{j_{i+1}=0\\ j_{i+1}\not\in X^\star_{i}}}^\np \cdots
    \sum_{\leftsubstack{j_n=0\\ j_n\not\in X^\star_{n-1}}}^\np
    \prod_{\leftsubstack{k=1\\ k\neq i}}^n \eta^\star_{k\comma j_k}
  }{
    \sum_{\leftsubstack{j_1=0\\ j_1\not\in X_0}}^\np
    \sum_{\leftsubstack{j_2=0\\ j_2\not\in X_1}}^\np \cdots
    \sum_{\leftsubstack{j_n=0\\ j_n\not\in X_{n-1}}}^\np
    \prod_{k=1}^n \eta_{k\comma j_k}
  }.
\end{equation}

The probability that a source $\Mp_j$ has no counterpart in $K$ is simply
given by
\begin{equation}
  \Poto(A_{0\comma j} \mid C \cap C') 
  = 
  1-\sum_{k=1}^n \Poto(A_{k\comma j}\mid C \cap C').
\end{equation}
\subsection{Likelihood and estimation of unknown parameters}
\label{fractionoto}
As in Sect.~\ref{fractionsto}, an estimate $\hat{\vec x}_\oto$ of the set
$\vec x$ of unknown parameters may be obtained by solving Eq.~\eqref{max_Lh}.
Under assumption~\ref{H_oto}, 
we obtain
from Eqs.~\eqref{P_oto(C|C')_res},
\eqref{Lh_gen}, and \eqref{lambda} that
\begin{equation}
  \label{Lh_oto_brut}
  \Lhoto 
  = 
  \Bigl(\sum_{\substack{j_1=0\\ j_1\not\in X_0}}^\np
  \sum_{\substack{j_2=0\\ j_2\not\in X_1}}^\np\cdots
  \sum_{\substack{j_n=0\\ j_n\not\in X_{n-1}}}^\np
  \prod_{k=1}^n \eta_{k\comma j_k}\Bigr)\multspace \prod_{k=1}^\np\xi_{0\comma k}.
\end{equation}
Because the number of terms in Eq.~\eqref{Lh_oto_brut} grows exponentially with
$n$ and $\np$, this equation seems useless.
In fact, the prior computation of $\Lhoto$ is not necessary if the probabilities
$\Poto(A_{i\comma j} \mid C \cap C')$ are calculable (we see how to evaluate
these in Sect.~\ref{impl_oto}).

Indeed, for any parameter $x_p$, we get the same result
(Eq.~\eqref{der(Lh_sto)/x}) as under assumption~\ref{H_sto}.
First, we note that, since the $\xi_{0\comma j}$ are independent of $\vec x$,
we obtain from Eq.~\eqref{Lh_gen} that
\begin{equation}
  \label{der(Lh)}
  \frac{\partial\ln\Lh}{\partial x_p} 
  =
  \frac{1}{\Prob(C \mid C')}\multspace
  \frac{\partial\Prob(C \mid C')}{\partial x_p}.
\end{equation}
Now, for any set $\Upsilon$ of indices and any product of strictly positive
functions $h_k$ of some variable $y$,
\begin{equation}
  \label{der(prod_g)}
  \frac{\partial\prod_{k\in \Upsilon} h_k}{\partial y}
  = 
  \sum_{\ell\in \Upsilon}{\frac{\partial h_\ell}{\partial y}\multspace
    \prod_{\substack{k\in \Upsilon\\ k\neq\ell}} h_k}
  = 
  \sum_{\ell\in \Upsilon}{\frac{\partial\ln h_\ell}{\partial y}\multspace
    \prod_{k\in \Upsilon} h_k}.
\end{equation}
With $h_k = \eta_{k\comma j_k}$, $y = x_p$ and $\Upsilon = \integinterv{1}{n}$,
we therefore obtain from Eq.~\eqref{P_oto(C|C')_res} that
\begin{equation}
  \label{der(P_oto)_gauche}
  \frac{\partial \Poto(C \mid C')}{\partial x_p}
  = 
  \Xi \multspace
  \sum_{\substack{j_1=0\\j_1\notin X_0}}^\np\sum_{\substack{j_2=0\\j_2\notin X_1}}^\np
  \cdots\sum_{\substack{j_n=0\\j_n\notin X_{n-1}}}^\np\sum_{i=1}^n{
    \frac{\partial\ln\eta_{i\comma j_i}}{\partial x_p}\multspace
    \prod_{k=1}^n\eta_{k\comma j_k}}.
\end{equation}
The expression of $\Poto(A_{i\comma j} \cap C \mid C')$ 
(Eq.~\eqref{P_oto(Aij,C|C')_res}) may also be written
\begin{equation}
  \Poto(A_{i\comma j} \cap C \mid C')
  = 
  \Xi \multspace
  \sum_{\substack{j_1=0\\j_1\notin X_0}}^\np\sum_{\substack{j_2=0\\j_2\notin X_1}}^\np
  \cdots\sum_{\substack{j_n=0\\j_n\notin X_{n-1}}}^\np
  {\chi(j_i = j)\multspace \prod_{k=1}^n \eta_{k\comma j_k}},
\end{equation}
where $\chi$ is the indicator function (i.e.\ $\chi(j_i = j) = 1$ if 
proposition ``$j_i = j$\kern1pt'' is true and $\chi(j_i = j) = 0$ otherwise),
so
\begin{align}
  \label{der(P_oto)_droite}
  \sum_{i=1}^n\sum_{j=0}^\np{\frac{\partial\ln\zeta_{i\comma j}}{\partial x_p}
    \multspace \Poto(A_{i\comma j} \cap C \mid C')}
  &=
  \Xi\multspace \sum_{i=1}^n\sum_{\substack{j_1=0\\j_1\notin X_0}}^\np
  \sum_{\substack{j_2=0\\j_2\notin X_1}}^\np
  \cdots\sum_{\substack{j_n=0\\j_n\notin X_{n-1}}}^\np\sum_{j=0}^\np
  {\chi(j_i = j)\multspace \frac{\partial\ln\zeta_{i\comma j}}{
      \partial x_p}\multspace
    \prod_{k=1}^n\eta_{k\comma j_k}}
  \notag \\
  &=
  \Xi\multspace \sum_{i=1}^n\sum_{\substack{j_1=0\\j_1\notin X_0}}^\np
  \sum_{\substack{j_2=0\\j_2\notin X_1}}^\np
  \cdots\sum_{\substack{j_n=0\\j_n\notin X_{n-1}}}^\np
  {\frac{\partial\ln\zeta_{i\comma j_i}}{\partial x_p}\multspace
    \prod_{k=1}^n\eta_{k\comma j_k}}.
\end{align}
If $j_i = 0$, then $\eta_{i\comma j_i} = \zeta_{i\comma j_i}$;
and if $j_i \neq 0$, the numerators of $\eta_{i\comma j_i}$ and $\zeta_{i\comma j_i}$ 
are the same and their denominators do not depend on $x_p$:
in all cases, 
$\partial\ln\eta_{i\comma j_i}/\partial x_p = 
\partial\ln\zeta_{i\comma j_i}/\partial x_p$.
The righthand sides of Eqs.~\eqref{der(P_oto)_gauche}
and \eqref{der(P_oto)_droite} are therefore identical.
Dividing their lefthand sides by $\Poto(C \mid C')$ and using 
Eqs.~\eqref{der(Lh)} and \eqref{P(Aij|C,C')_gen}, 
we obtain, as announced,
\begin{equation}
  \label{der(Lh_oto)/x}
  \frac{\partial\ln\Lhoto}{\partial x_p}
  = 
  \sum_{i=1}^n\sum_{j=0}^\np{\frac{\partial\ln\zeta_{i\comma j}}{
      \partial x_p}\multspace \Poto(A_{i\comma j} \mid C \cap C')}.
\end{equation}

For $x_p = f$ in particular, because of Eq.~\eqref{somme_j},
and as  under assumption~\ref{H_sto}, Eq.~\eqref{der(Lh_oto)/x} reduces to
\begin{equation}
  \label{der(Lh_oto)/f}
  \frac{\partial\ln\Lhoto}{\partial f}
  = \frac{
    n\multspace (1-f) - \sum_{i=1}^n \Poto(A_{i\comma 0} \mid C \cap C')
  }{
    f\multspace (1-f)
  }.
\end{equation}
From Eq.~\eqref{max_Lh}, a maximum likelihood estimator of $f$ is thus
\begin{equation}
  \hat f_\oto 
  = 
  1 - \frac{1}{n}\multspace
  \sum_{i=1}^n\expandafter\hat\Poto(A_{i\comma 0} \mid C \cap C'),
  \label{f_est_oto}
\end{equation}
where $\expandafter\hat\Poto$ is the value of $\Poto$ at $f = \hat f_\oto$.

To compare assumptions \ref{H_sto}, \ref{H_oto}, and \ref{H_sto} and to select
the most appropriate one to compute $\Prob(A_{i\comma j} \mid C \cap C')$,
an expression is needed for $\Lhoto$.
If probabilities $\Poto(A_{i\comma 0} \mid C \cap C')$ are
calculable, $\Lhoto$ may be obtained for any $f$ by integrating
Eq.~\eqref{der(Lh_oto)/f} with respect to $f$.
Since all $K$-~and $K'$-sources are unrelated and randomly
distributed for $f = 0$, the integration constant is
(cf.~Eq.~\eqref{Lh_oto_brut})
\begin{equation}
  \label{Lh_oto(0)}
  {\bigl(\ln\Lhoto\bigr)}_{f=0}
  = 
  \sum_{i=1}^n \ln\xi_{i\comma0} + \sum_{j=1}^\np \ln\xi_{0\comma j}.
\end{equation}
\section{Practical implementation: the \Aspects\ code}
\label{code}
\subsection{Overview}
{\emergencystretch=1em
To implement the results established in Sects.~\ref{global}, \ref{fractionsto},
\ref{sect_P_oto}, and \ref{fractionoto}, we have built a Fortran~95 code, 
\Aspects\ --~a French acronym (pronounced [asp$\varepsilonup$] in International 
Phonetic Alphabet, not [{\ae}spekts]) for
``\!\emph{\textsc{\textbf{As}}sociation
  \textsc{\textbf{p}}ositionnell\textsc{\textbf{e}}\slash
  \textsc{\textbf{p}}robabilist\textsc{\textbf{e}}
  de \textsc{\textbf{c}}a\textsc{\textbf{t}}alogues
  de \textsc{\textbf{s}}ources}'', 
or ``probabilistic positional association of source catalogs'' in English.
The source files are freely available%
\footnote{%
  Fortran~90 routines from \emph{Numerical Recipes} \citep{NR}
  are used to sort arrays and locate a value in an ordered table.
  Because of license constraints, we cannot provide them,
  but they may easily be replaced by free equivalents.
  
  \emph{\textbf{Corrigendum}: \emph{Numerical Recipes} routines are not needed anymore.}
}
at \href{http://www2.iap.fr/users/fioc/Aspects/}{www2.iap.fr/users/fioc/Aspects/}.
The code compiles with \textsc{IFort} and \textsc{GFortran}.
\par
}

Given two catalogs of sources with their positions and the uncertainties on 
these, \Aspects\ computes, under assumptions~\ref{H_sto}, \ref{H_oto}, and 
\ref{H_ots}, the overall likelihood $L$, estimates of $f$ and $f'$,
and the probabilities $\Prob(A_{i\comma j} \mid C \cap C')$.
It may also simulate all-sky catalogs for various association models
(cf.~Sect.~\ref{mock}).

We provide hereafter explanations of general interest for the practical 
implementation in \Aspects\ of Eqs.~\eqref{P_sto(Aij|C,C')_res1}, 
\eqref{f_est_sto1}, \eqref{Lh_sto}, \eqref{P_oto(Aij|C,C')_res}, 
\eqref{f_est_oto}, and \eqref{Lh_oto_brut}.
Some more technical points (such as the procedures used to search for nearby
objects, simulate the positions of associated sources and integrate
Eq.~\eqref{der(Lh_oto)/f}) are only addressed in appendices
to the documentation of the code \citep{doc_Aspects}.
The latter also contains the following complements:
another (but equivalent) expression for $\Lhoto$,
formulae derived under $H_\ots$,
computations under $H_\oto$ for $n > \np$,
a calculation of the uncertainties on unknown parameters under $H_\sto$,
and a proof of Eq.~\eqref{concave}.
\subsection{Elimination of unlikely counterparts}
\label{vicinity}
Under assumption~\ref{H_sto}, computing the probability of association
$\Psto(A_{i\comma j} \mid C \cap C')$ between $M_i$ and $\Mp_j$
from Eq.~\eqref{P_sto(Aij|C,C')_res1}
is straightforward if $f$ and the positional uncertainties are known.
However, the number of calculations for the whole sample
or for determining $\hat{\vec x}$ is on the order of
$n\multspace \np$, a huge number for the catalogs available nowadays.
We must therefore try to eliminate all unnecessary computations.

Since $\xi_{i\comma k}$ is given by a normal law if $i \neq 0$ and $k \neq 0$, it
rapidly drops to almost $0$ when 
we consider sources $\Mp_k$ at increasing angular distance $\psi_{i\comma k}$
from $M_i$.
Therefore, 
there is no need to compute $\Psto(A_{i\comma j} \mid C \cap C')$
for all couples $(M_i, \Mp_j)$ or to sum on all $k$ from $1$ to $\np$ 
in Eq.~\eqref{P_sto(Aij|C,C')_res2}.
More explicitly, let $R'$ be some angular distance 
such that, for all $(M_i, \Mp_k)$, if $\psi_{i\comma k} \geqslant R'$
then $\xi_{i\comma k} \approx 0$,
say
\begin{equation}
  \label{def_R'}
  R' \ga 5\multspace
  \!\sqrt{
    \smash[b]{
      \max_{\ell\in\integinterv{1}{n}} a_\ell^2 + \max_{\ell\in\integinterv{1}{\np}} a_{\ell}'^2
    }
  },
  \vphantom{
    \max_{\ell\in\integinterv{1}{n}} a_\ell^2 + \max_{\ell\in\integinterv{1}{\np}} a_{\ell}'^2
  }
\end{equation}
where the $a_\ell$ and $a'_{\smash[t]{\ell}}$ are the semi-major axes
of the positional uncertainty ellipses of $K$-~and $K'$-sources
(cf.~App.~\ref{identical,known};
the square root in Eq.~\eqref{def_R'} is thus the maximal possible
uncertainty on the relative position of associated sources).
We may set $\Psto(A_{i\comma j} \mid C \cap C')$ to $0$ if $\psi_{i\comma j} > R'$,
and replace the sums 
$\smash[t]{\sum_{\smash[t]{k=1}}^\np}$ by 
$\smash[t]{\sum_{\smash[t]{k=1{;}\, \psi_{i\comma k}\leqslant R'}}^\np}$ 
in Eq.~\eqref{P_sto(Aij|C,C')_res2}:
only nearby $K'$-sources matter.
\subsection{Fraction of sources with a counterpart}
\label{cvg_f}
All the probabilities depend on $f$ and, possibly, on other unknown parameters
like $\sigmatot$ and $\nutot$ (cf.~Apps.~\ref{unknown} and \ref{different}).
Under assumption~\ref{H_sto}, estimates of these parameters may be found 
by solving Eq.~\eqref{max_Lh} using Eq.~\eqref{der(Lh_sto)/x}.

If the fraction of sources with a counterpart is the only unknown,
the $\xi_{i\comma j}$ need to be computed only once and $\hat f_\sto$ may easily 
be determined from Eq.~\eqref{f_est_sto1} by an iterative procedure.
Denoting by $g$ the function
\begin{equation}
  \label{fonction_g}
  g\colon f \in [0, 1] \longmapsto
  1-\frac{1}{n}\multspace \sum_{i=1}^n\Psto(A_{i\comma 0} \mid C \cap C'),
\end{equation}
we now prove 
that, for any $f_0 \in \mathopen]0, 1\mathclose[$, the sequence
$(f_k)_{k\in\varmathbb{N}}$ defined by $f_{k+1} \coloneqq g(f_k)$ tends to
$\hat f_\sto$.

As is obvious from Eq.~(\ref{P_sto(Aij|C,C')_res2}b), 
$\Psto(A_{i\comma 0} \mid C \cap C')$ decreases for all $i$ when $f$ increases:
$g$ is consequently an increasing function.
Note also that, from Eqs.~\eqref{der(Lh_sto)/f} and \eqref{fonction_g},
\begin{equation}
  \label{expression_g}
  g(f) 
  = 
  f + \frac{f\multspace (1-f)}n \multspace \frac{\partial\ln\Lhsto}{\partial f}.
\end{equation}
The only fixed points of $g$ are thus $0$, $1$ and the \vadjust{\vspace{-0.5pt}}unique solution 
$\hat f_\sto$ to $\partial\ln\Lhsto/\partial f = 0$.
Because $\partial^2\ln\Lhsto/\partial f^2 < 0$ (cf.~Eq.~\eqref{concave})
and $\hat\partial\ln\Lhsto/\hat\partial f = 0$, we have
$\partial\ln\Lhsto/\partial f \geqslant 0$ if $f \in [0, \hat f_\sto]$,
so $g(f) \geqslant f$ in this interval by Eq.~\eqref{expression_g}.
Similarly, if $f \in [\hat f_\sto, 1]$,
then
$\partial\ln\Lhsto/\partial f \leqslant 0$ and thus $g(f) \leqslant f$.

Consider the case $f_0 \in \mathopen{]}0, \hat f_\sto]$.
If $f_k \leqslant \hat f_\sto$, then as just shown, $g(f_k) \geqslant f_k$;
we also have
$g(f_k) \leqslant g(\hat f_\sto) = \hat f_\sto$, \vadjust{\vspace{-0.5pt}}because $g$ is an 
increasing function and $\hat f_\sto$ is a fixed point of it.
Since $g(f_k) = f_{k+1}$, the sequence $(f_k)_{k\in\varmathbb{N}}$ is
increasing and \vadjust{\vspace{-0.5pt}}bounded from above by $\hat f_\sto$:
it therefore converges in $[f_0, \hat f_\sto]$.
Because $g$ is continuous and $\hat f_\sto$ is the only fixed point in this
interval, $(f_k)_{k\in\varmathbb{N}}$ tends to $\hat f_\sto$.
Similarly, if $f_0 \in [\hat f_\sto, 1\mathclose{[}$, then
$(f_k)_{k\in\varmathbb{N}}$ is a decreasing sequence converging to $\hat f_\sto$.

Because of Eq.~\eqref{f_est_oto}, this procedure also works in practice
under assumption~\ref{H_oto} (with $\Psto$ replaced by $\Poto$ in 
Eq.~\eqref{fonction_g}), although it is not obvious that
$\Poto(A_{i\comma 0} \mid C \cap C')$ decreases for all $i$ when $f$ increases, 
nor that $\partial^2\ln\Lhoto/\partial f^2 < 0$.
A good starting value $f_0$ may be $\hat f_\sto$.
\subsection{Computation of one-to-one probabilities of association}
\label{impl_oto}
What was said in Sect.~\ref{vicinity} about eliminating unlikely 
counterparts in the calculation of probabilities under $H_\sto$ still holds 
under $H_\oto$.
However, because of the combinatorial explosion of the number of terms in 
Eq.~\eqref{P_oto(Aij|C,C')_res}, computing
$\Poto(A_{i\comma j} \mid C \cap C')$ exactly is still clearly hopeless.
Yet, after some wandering (Sects.~\ref{first_try} and \ref{failure}),
we found a working solution (Sect.~\ref{working}).
\subsubsection{A first try}
\label{first_try}
Our first try was inspired by the (partially wrong) idea that, although all 
$K$-sources are involved in the numerator and denominator of 
Eq.~\eqref{P_oto(Aij|C,C')_res},
only those close to $M_i$ should matter in their \emph{ratio}.
A sequence of approximations converging to the true value of
$\Poto(A_{i\comma j} \mid C \cap C')$ might then be built as follows
(all quantities defined or produced in this first try are written with the superscript
``\wrong'' for ``wrong'').

To make things clear, consider $M_1$ and some possible counterpart
$\Mp_j$ within its neighborhood ($\psi_{1\comma j} \leqslant R'$)
and assume that $M_2$ is the first nearest neighbor of $M_1$ in $K$,
$M_3$ its second nearest neighbor, etc.
For any $d \in \integinterv{1}{n}$, define
\begin{equation}
  p^\wrong_{\smash[t]{d}}(1, j) 
  \coloneqq
  \frac{
    \zeta_{1\comma j}\multspace \sum_{\leftsubstack{j_2=0\\ j_2\not\in X^{\star}_1}}^\np
    \cdots \sum_{\leftsubstack{j_d=0\\ j_d\not\in X^{\star}_{d-1}}}^\np
    \prod_{k=2}^d \eta^{\star}_{k\comma j_k}
  }{
    \sum_{\leftsubstack{j_1=0\\ j_1\not\in X_0}}^\np \sum_{\leftsubstack{j_2=0\\ j_2\not\in X_1}}^\np
    \cdots \sum_{\leftsubstack{j_d=0\\ j_d\not\in X_{d-1}}}^\np \prod_{k=1}^d \eta_{k\comma j_k}
  }.
\end{equation}
The quantity $p^\wrong_{\smash[t]{d}}(1, j)$ thus depends only on $M_1$ and its $d-1$
nearest neighbors in $K$.
As $p^\wrong_n(1, j)$ is the \emph{one-to-one} probability of association between
$M_1$ and $\Mp_j$ (cf.~Eq.~\eqref{P_oto(Aij|C,C')_res}), the sequence 
$(p^\wrong_{\smash[t]{d}}[1, j])$ tends to $\Poto(A_{1\comma j} \mid C \cap C')$
when the depth $d$ of the recursive sums tends to $n$.
After some initial fluctuations, $p^\wrong_{\smash[t]{d}}(1, j)$ enters a steady 
state.
This occurs when $\psi(M_1, M_{d+1})$ exceeds a distance $R$ equal to
a few times $R'$ (at least $2\multspace R'$).
We may therefore think that the convergence is then achieved
and stop the recursion at this $d$.
It is all the more tempting that 
$p^\wrong_1(1, j) = \Psto(A_{1\comma j} \mid C \cap C')$
and that the several-to-one probability looks like a first-order approximation
to $\Poto$\textellipsis

More formally and generally, for any $M_i$, let $\phi$ be a permutation on $K$
ordering the elements $M_{\phi(1)}$, $M_{\phi(2)}$,~\textellipsis, $M_{\phi(n)}$
by increasing angular distance to $M_i$ (in particular, $M_{\phi(1)} = M_i$).
For $j=0$ or $\Mp_j$ within a distance $R'$ (cf.~Sect.~\ref{vicinity}) from 
$M_i$, and for any $d \in \integinterv{1}{n}$, define
\begin{equation}
  \label{P_oto_iter_w}
  p^\wrong_{\smash[t]{d}}(i, j)
  \coloneqq
  \frac{
    \zeta_{i\comma j}\multspace\sum_{\leftsubstack{j_2=0\\ j_2\not\in \widetilde X^{\star}_1}}^\np
    \cdots \sum_{\leftsubstack{j_d=0\\ j_d\not\in \widetilde X^{\star}_{d-1}}}^\np
    \prod_{k=2}^d \widetilde\eta^{\,\star\,\wrong}_{k\comma j_k}
  }{
    \sum_{\leftsubstack{j_1=0\\ j_1\not\in \widetilde X_0}}^\np
    \sum_{\leftsubstack{j_2=0\\ j_2\not\in \widetilde X_1}}^\np \cdots
    \sum_{\leftsubstack{j_d=0\\ j_d\not\in \widetilde X_{d-1}}}^\np
    \prod_{k=1}^d \widetilde\eta^{\,\wrong}_{k\comma j_k}
  },
\end{equation}
where, as in Eqs.~\eqref{def_J}, \eqref{def_J*}, \eqref{def_eta}, and 
\eqref{def_eta*},
\begin{gather}
  \widetilde X_k \coloneqq X_k\quad  \text{for all } k \in \integinterv{0}{n};
  \qquad
  \widetilde X^{\star}_1 \coloneqq \{j\} \setminus \{0\};
  \qquad
  \widetilde X^{\star}_k 
  \coloneqq 
  (\widetilde X^{\star}_{k-1} \cup \{j_k\}) \setminus \{0\}
  \quad\text{for all } k \in \integinterv{2}{n};
  \\
  \label{tilde}
  \widetilde\eta^{\,\wrong}_{k\comma 0} \coloneqq
  \widetilde\eta^{\,\star\,\wrong}_{k\comma 0} \coloneqq \zeta_{\phi(k)\comma 0};
  \qquad
  \widetilde\eta^{\,\wrong}_{k\comma j_k}
  \coloneqq \frac{f\multspace \xi_{\phi(k)\comma j_k}}{\np-\card \widetilde X_{k-1}}
  \quad\text{and}\quad
  \widetilde\eta^{\,\star\,\wrong}_{k\comma j_k}
  \coloneqq 
  \frac{f\multspace \xi_{\phi(k)\comma j_k}}{\np-\card \widetilde X^{\star}_{k-1}}
  \quad \text{for } j_k \neq 0.
\end{gather}
Let
\begin{equation}
  \label{min_prof}
  \depthmin(i) 
  \coloneqq 
  \min\Bigl(d \in \integinterv{1}{n} \bigm| \psi[M_i, M_{\phi(d+1)}] > R\Bigr).
\end{equation}
Given above considerations, $\Poto(A_{i\comma j} \mid C \cap C')$ can be 
evaluated as $p^\wrong_\oto(i, j) \coloneqq p^\wrong_{\smash[t]{\depthmin(i)}}(i, j)$.

The computation of $p^\wrong_{\smash[t]{d}}(i, j)$ may be further restricted
(and in practice, because of the recursive sums in Eq.~\eqref{P_oto_iter_w},
\emph{must be}) to sources $\Mp_{j_k}$ in the neighborhood of the objects 
$(M_{\phi(k)})_{k\in\integinterv{1}{d}}$, as explained in Sect.~\ref{vicinity}.
\subsubsection{Failure of the first try}
\label{failure}
To test the reliability of the evaluation of $\Poto(A_{i\comma j} \mid C \cap C')$
by $p^\wrong_\oto(i, j)$, \vadjust{\vspace{-1pt}}we simulated all-sky mock catalogs for one-to-one
associations and analyzed them with a first version of \Aspects.
Simulations were run for $f = 1/2$, $\np = \cramped{10^5}$, 
$n \in \integinterv{\cramped{10^3}}{\cramped{10^5}}$,
and known circular positional uncertainties with $\sigmatot = 10^{-3}\,\radian$
(see Sects.~\ref{mock} and \ref{simul_f} for a detailed description).

Three estimators of $f$ were compared to the input value:
\begin{itemize}
\item
  $\hat f_\sto$, the value maximizing $\Lhsto$ (Eq.~\eqref{f_est_sto1});
\item
  $\hat f_\oto^\wrong$, the value maximizing the one-to-one likelihood 
  $\Lhoto^\wrong$ derived from the $p^\wrong_\oto$.
  This estimator is computed from Eq.~\eqref{f_est_oto} with $p^\wrong_\oto(i, 0)$
  instead of $\Poto(A_{i\comma 0} \mid C \cap C')$;
\item
  $\hat f_\ots$, an estimator built from the one-to-several assumption in the
  following way: because \ref{H_ots} is \vadjust{\vspace{-0.5pt}}fully symmetric to \ref{H_sto}, 
  we just need
  to swap $K$ and $K'$ (i.e., swap $f$ and $f'$, $n$ and $\np$, etc.)\ 
  in Eqs.~\eqref{P_sto(Aij|C,C')_res2}, \eqref{P_sto(A0j|C,C')}, and 
  \eqref{f_est_sto1} to \vadjust{\vspace{-1pt}}obtain $\hat f'_\ots$ instead of $\hat f_\sto$,
  and then, from Eq.~\eqref{f'_est_sto}, $\hat f_\ots$ instead of $\hat f'_\sto$.
  The one-to-several likelihood $\Lhots$ is computed from Eq.~\eqref{Lh_sto} 
  in the same way.
\end{itemize}
\begin{figure}
  \resizebox{\hsize}{!}{\includegraphics{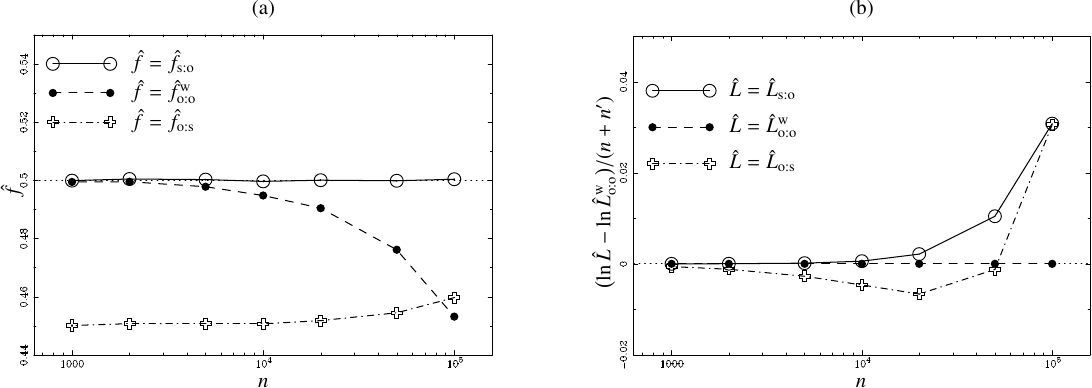}}
  \caption{%
    One-to-one simulations for $f = 1/2$, $\np = 10^5$, and circular positional 
    uncertainty ellipses with $\sigmatot = 10^{-3}\,\radian$ (see 
    Sects.~\ref{mock} and \ref{simul_f} for details).%
    \enskip
    \textbf{(a)}~Mean value of different estimators $\hat f$ of $f$ as a 
    function of $n$.
    The dotted line indicates the input value of $f$.%
    \enskip
    \textbf{(b)}~Normalized average maximum value $\hat\Lh$ of different 
    likelihoods as a function of $n$, compared to
    $\expandafter\hat\Lhoto^\wrong$.
  }
  \label{fig1}
\end{figure}
The mean values of these estimators are plotted as a function of $n$
in Fig.~\ref{fig1}a \vadjust{\vspace{-1pt}}(error bars are smaller than the size of the points).
As is obvious, the \emph{ad~hoc} estimator $\hat f_\oto^\wrong$ diverges 
from $f$ when $n$ increases.
This statistical inconsistency%
\footnote{%
  A consistent estimator is a statistic converging to the true value of a 
  parameter when the size of the sample from which it is derived increases.
  The concept of consistency is not very clear in the context
  of this paper, since there are two sample sizes, $n$ and $\np$.
}
seems surprising for a maximum likelihood estimator
since the model on which it is based is correct by construction.
However, all the demonstrations of consistency of maximum likelihood
estimators we found in the \vadjust{\vspace{-1pt}}literature 
(e.g., in \citealp{KS})
rest on the assumption that the overall likelihood is the
\emph{product} of the probabilities of each datum,
which is not the case for $\Lhoto$ (cf.~Eq.~\eqref{Lh_oto_brut}).
Since $\hat f_\sto$ is a good estimator of $f$, it might be used
to compute $\Poto(A_{i\comma j} \mid C \cap C')$ from
$p^\wrong_\oto(i, j)$ --~if the latter correctly \vadjust{\vspace{-1.5pt}}approximates the former.
By itself, the inconsistency of $\hat f^\wrong_\oto$ is therefore not a problem.

More embarrassing is that \ref{H_oto} is not the most likely 
assumption (see Fig.~\ref{fig1}b):
the mean value of $\expandafter\hat\Lhoto^\wrong$ is less than that of 
$\expandafter\hat\Lhsto$ over the full interval of $n$\,!
These two failures hint that the sequence $(p^\wrong_{\smash[t]{d}}[i, j])$ has
not yet converged to $\Poto(A_{i\comma j} \mid C \cap C')$ at $d = \depthmin(i)$.

To check this, we ran simulations with small numbers of sources ($n$ and
$\np$ less than $10$), so that we could compute $p^\wrong_n(i, j)$ exactly and
study how $(p^\wrong_{\smash[b]d}[i, j])$ tends to it.
To test whether source \vadjust{\vspace{-1.5pt}}confusion might be the reason for the problem,
we created mock catalogs with very large positional uncertainties%
\footnote{%
  Small positional uncertainties could also be used if sources were
  distributed on a small fraction of the sky, but there might be side effects.
}
$\sigmatot$, comparable to the distance between unrelated sources.
Because the expressions given in App.~\ref{cov} for $\xi_{i\comma j}$ are for planar
normal laws and become wrong
when the \vadjust{\vspace{-1pt}}distance between $M_i$ and $\Mp_j$ is more than a few degrees
because of the curvature,
we ran simulations on a whole circle instead of a sphere;
nevertheless, we took $\sigmatot \la 30^\circ$ because the linear normal law is 
inappropriate on a circle for higher values, due to its finite extent.
What we found is that, after the transient phase where it oscillates,
$(p^\wrong_{\smash[t]{d}}[i, j])$ slowly drifts to 
$\Poto(A_{i\comma j} \mid C \cap C')$ and only converges at $d = n$\,!
This drift was imperceptible for the high values of $n$ and $\np$
used in Sect.~\ref{first_try}.
\subsubsection{Reconsideration and solution}
\label{working}
To understand where the problem comes from, we consider the simplest case
of interest: $n = \np = 2$.
We assume moreover that $\xi_{1\comma 2} \approx \xi_{2\comma 1} \approx 0$.
We then have
\begin{align}
  \Poto(C \mid C')
  &\approx
  \Biggl([1-f]^2\multspace \xi_{1\comma 0}\multspace \xi_{2\comma 0}
  + \frac{[1-f]\multspace f}{2}\multspace [\xi_{1\comma 0}\multspace \xi_{2\comma 2}
  + \xi_{1\comma 1}\multspace \xi_{2\comma 0}]
  +\frac{f^2}{2}\multspace \xi_{1\comma 1}\multspace \xi_{2\comma 2}\Biggr)
  \multspace \df^2\vec r_1\multspace \df^2\vec r_2,
  \\
  \Poto(A_{1\comma 0} \cap C \mid C')
  &\approx
  (1-f)\multspace \xi_{1\comma 0}\multspace
  \Biggl([1-f]\multspace \xi_{2\comma 0} + \frac{f}{2}\multspace 
  \xi_{2\comma 2}\Biggr) \multspace \df^2\vec r_1\multspace \df^2\vec r_2,
  \\
  \Poto(A_{1\comma 1} \cap C \mid C')
  &\approx
  \frac{f}{2}\multspace \xi_{1\comma 1}\multspace \Bigl([1-f]\multspace 
  \xi_{2\comma 0} + f\multspace \xi_{2\comma 2}\Bigr)\multspace \df^2\vec r_1
  \multspace \df^2\vec r_2.
\end{align}
The probabilities 
$\Poto(A_{1\comma j} \mid C \cap C') =
\Poto(A_{1\comma j} \cap C \mid C')/\Poto(C \mid C')$
obviously depend on $\xi_{2\comma 2}$.
In particular,
\begin{equation}
  \label{<<}
  \text{if } \xi_{2\comma 2} \ll \xi_{2\comma 0},
  \qquad
  \Poto(A_{1\comma 0} \mid C \cap C') \approx
  \frac{(1-f)\multspace \xi_{1\comma 0}}{
    (1-f)\multspace \xi_{1\comma 0} + f\multspace \xi_{1\comma 1}/2}
  \quad\text{and}\quad
  \Poto(A_{1\comma 1} \mid C \cap C') \approx
  \frac{f\multspace \xi_{1\comma 1}/2}{
    (1-f)\multspace \xi_{1\comma 0} + f\multspace \xi_{1\comma 1}/2};
\end{equation}
in that case, $\Poto(A_{2\comma 2} \mid C \cap C') \approx 0$,
and both $\Mp_1$ and $\Mp_2$ are free for $M_1$.
On the other hand,
\begin{equation}
  \label{>>}
  \text{if } \xi_{2\comma 2} \gg \xi_{2\comma 0},
  \qquad
  \Poto(A_{1\comma 0} \mid C \cap C') \approx
  \frac{(1-f)\multspace \xi_{1\comma 0}}{
    (1-f)\multspace \xi_{1\comma 0} + f\multspace \xi_{1\comma 1}/1}
  \quad\text{and}\quad
  \Poto(A_{1\comma 1} \mid C \cap C') \approx
  \frac{f\multspace \xi_{1\comma 1}/1}{
    (1-f)\multspace \xi_{1\comma 0} + f\multspace \xi_{1\comma 1}/1};
\end{equation}
in that case, $\Poto(A_{2\comma 2} \mid C \cap C') \approx 1$:
$M_2$ and $\Mp_{2}$ are almost certainly bound, so $\Mp_{2}$ may not be
associated to $M_1$, and $\Mp_{1}$ is the only possible counterpart of $M_1$.

The difference between the results obtained for
$\xi_{2\comma 2} \ll \xi_{2\comma 0}$ and $\xi_{2\comma 2} \gg \xi_{2\comma 0}$
shows that probabilities $\Poto(A_{1\comma j} \mid C \cap C')$
depend on the relative positions of $M_2$ and $\Mp_{2}$, even
when both $M_2$ and $\Mp_{2}$ are distant from $M_1$ and $\Mp_{1}$:
unlike the idea stated in Sect.~\ref{first_try},
distant $K$-sources do matter for $\Poto$ probabilities!
However, as highlighted by the ``$/2$'' and ``$/1$'' factors in
Eqs.~\eqref{<<} and \eqref{>>}, the distant $K$-source $M_2$ only changes the
\emph{number} of $K'$-sources (two for $\xi_{2\comma 2} \ll \xi_{2\comma 0}$,
one for $\xi_{2\comma 2} \gg \xi_{2\comma 0}$) that may be identified to $M_1$:
its exact position is unimportant.

This suggests the following solution:
replace $\np$ in Eq.~\eqref{tilde} by the number $\npeff(i,d)$ of 
$K'$-sources that may effectively be associated to $M_i$ and its $d-1$ nearest 
neighbors in $K$;
i.e., dropping the superscript ``\wrong'', define
\begin{equation}
  \label{P_oto_iter}
  p_d(i, j)
  \coloneqq
  \frac{\zeta_{i\comma j}\multspace
    \sum_{\leftsubstack{j_2=0\\ j_2\not\in \widetilde X^{\star}_1}}^\np
    \cdots \sum_{\leftsubstack{j_d=0\\ j_d\not\in \widetilde X^{\star}_{d-1}}}^\np
    \prod_{k=2}^d \widetilde\eta^{\,\star}_{k\comma j_k}
  }{
    \sum_{\leftsubstack{j_1=0\\ j_1\not\in \widetilde X_0}}^\np
    \sum_{\leftsubstack{j_2=0\\ j_2\not\in \widetilde X_1}}^\np
    \cdots
    \sum_{\leftsubstack{j_d=0\\ j_d\not\in \widetilde X_{d-1}}}^\np
    \prod_{k=1}^d \widetilde\eta_{k\comma j_k}
  },
\end{equation}
where
\begin{equation}
  \widetilde\eta_{k\comma 0} \coloneqq
  \widetilde\eta^{\,\star}_{k\comma 0} \coloneqq \zeta_{\phi(k)\comma 0};
  \qquad
  \widetilde\eta^{\,\star}_{k\comma j_k}
  \coloneqq \frac{f\multspace \xi_{\phi(k)\comma j_k}}{
    \npeff(i{,}\,d)-\card \widetilde X^{\star}_{k-1}}
  \quad\text{and}\quad
  \widetilde\eta_{k\comma j_k}
  \coloneqq \frac{f\multspace \xi_{\phi(k)\comma j_k}}{
    \npeff(i{,}\,d)-\card \widetilde X_{k-1}}
  \quad\text{for } j_k \neq 0,
\end{equation}
and use $p_\oto(i, j) \coloneqq p_{\smash[t]{\depthmin(i)}}(i, j)$, where 
$\depthmin(i)$ is defined by Eq.~\eqref{min_prof}, to evaluate
$\Poto(A_{i\comma j} \mid C \cap C')$.

An estimate of $\npeff$ is given by%
\footnote{%
  Equation~\eqref{n'_eff} is valid for any $f \in [0, 1]$.
  When $f \approx \hat f_\oto$, it is more efficient to make the approximation
  $\npeff(i,d) \approx \np -f\multspace (n-d)$:
  \vadjust{\vspace{-1pt}}this expression accelerates the convergence to $\hat f_\oto$ of the
  sequence $(f_k)$ defined in Sect.~\ref{cvg_f}.
}
\begin{equation}
  \label{n'_eff}
  \npeff(i,d) =
  \np - \sum_{k=d+1}^n{\Bigl(1 - \Poto[A_{\phi(k)\comma0} \mid C \cap C']\Bigr)}.
\end{equation}
The sum in Eq.~\eqref{n'_eff} is nothing but the typical number of counterparts
in $K'$ associated to distant $K$-sources.
Note that $\npeff(i, d = n) = \np$, so 
we recover the theoretical result for $\Poto(A_{i\comma j} \mid C \cap C')$ when all sources are considered.
As $\Poto$ depends on $\npeff$ which in turn depends on $\Poto$,
both may be computed with a back and forth iteration;
this procedure converges in a few steps
if, instead of $\Poto$, the value of $\Psto$ is taken to initiate the sequence.
\subsection{Tests of \Aspects}
\label{checks}
As computations made under assumption~\ref{H_oto} are complex (they involve 
recursive sums for instance), we made several consistency checks of the code.
In particular, we \vadjust{\vspace{-1pt}}swapped $K$ and $K'$ for $n \neq \np$ and compared 
quantities resulting from this swap (written with the superscript ``$\leftrightarrow$'')
to original ones:
within numerical errors, $\hat f'^\leftrightarrow_\oto = \hat f_\oto$ and,
for $f'^\leftrightarrow = f$, we get $\Lhoto^\leftrightarrow = \Lhoto$ and
$\Poto^\leftrightarrow(A_{j\comma i} \mid C' \cap C) =
\Poto(A_{i\comma j} \mid C \cap C')$ for all $(M_i, \Mp_j)$.

We moreover numerically checked for small $n$ and $\np$ ($\la 5$) that
Eq.~\eqref{Lh_oto_brut} and the integral of Eq.~\eqref{der(Lh_oto)/f}
with respect to $f$ are consistent and that \Aspects\ returns the same value 
as \textsc{Mathematica} \citep{Mathematica}.
For even smaller $n$ and $\np$ ($\leqslant 3$), we confirmed
that manual analytical expressions, obtained from the enumeration
of all possible associations between $K$ and $K'$, are identical to 
\textsc{Mathematica}'s symbolic calculations.
For the large $n$ and $\np$ of practical interest, although we did not give a 
formal proof of the solution of Sect.~\ref{working}, the analysis of 
simulations (Sect.~\ref{simul}) makes us confident in the code.
\section{Simulations}
\label{simul}
In this section, we analyze various estimators of the unknown parameters.
Because of the complexity of the expressions we obtained,
we did not try to do it analytically but used simulations.
We also compare the likelihood of the assumptions \ref{H_sto},
\ref{H_oto}, and \ref{H_ots}, given the data.
\subsection{Creation of mock catalogs}
\label{mock}
We have built all-sky mock catalogs with \Aspects\
in the cases of several-~and one-to-one associations.
To do this, we first selected the indices of $f\multspace n$ objects in $K$,
and associated randomly the index of a counterpart in $K'$ to each of them;
for one-to-one simulations, a given $K'$-source was associated at most once.
We then drew the true positions of $K'$-sources uniformly on the sky.
The true positions of $K$-sources without counterpart were also drawn
in the same way;
for sources with a counterpart, we took the true position of their counterpart.
The observed positions of $K$-~and $K'$-sources were finally computed
from the true positions for given parameters $(a_i, b_i, \beta_i)$
and $(a'_{\smash[t]{j}}, b'_{\smash[t]{j}}, \betapj)$ of the positional uncertainty
ellipses (see App.~\ref{identical,known}).
\subsection{Estimation of \texorpdfstring{$f$}{\$f\$} if positional uncertainty ellipses are known and circular}
\label{simul_f}
\begin{figure}
  \resizebox{\hsize}{!}{\includegraphics{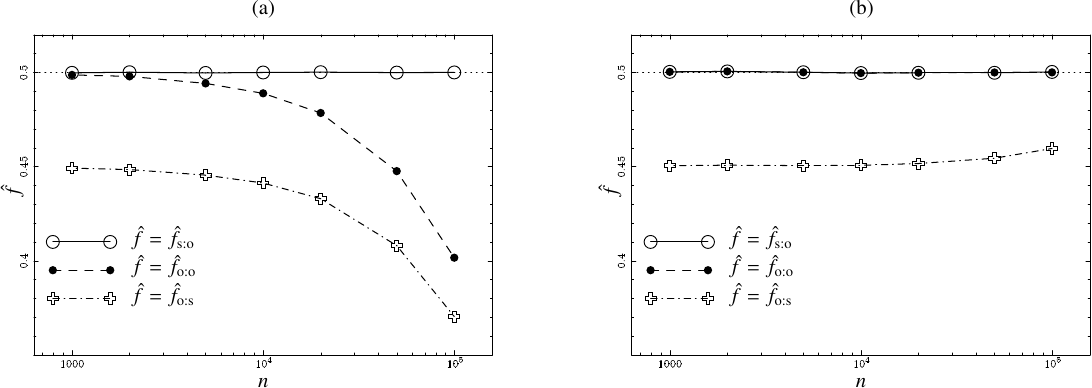}}
  \caption{%
    Mean value of different estimators $\hat f$ of $f$ as a function of $n$
    for $f = 1/2$ (dotted line), $\np = 10^5$, and circular positional 
    uncertainty ellipses with $\sigmatot = 10^{-3}\,\radian$
    (see Sects.~\ref{mock} and \ref{simul_f} for details).%
    \enskip \textbf{(a)}~Several-to-one simulations.%
    \enskip \textbf{(b)}~One-to-one simulations
    ($\hat f_\sto$ and $\hat f_\oto$ overlap).
  }
  \label{fig2}
\end{figure}

Mock catalogs were created with $a_i = b_i = \sigma$ (see notations in 
App.~\ref{identical,known}) for all $M_i \in K$ and with
$a'_{\smash[t]{j}} = b'_{\smash[t]{j}} = \sigma'$ for all $\Mp_j \in K'$.
Positional uncertainty ellipses are therefore circular here.
Only two parameters matter in that case: $f$ and
\begin{equation}
  \sigmatot \coloneqq \!\sqrt{\sigma^2+\sigma'^2}.
\end{equation}
Hundreds of simulations were run for $f = 1/2$, $\np = 10^5$,
$\sigmatot = 10^{-3}\,\radian$, and $n \in \integinterv{10^3}{10^5}$.
We analyzed them with \Aspects, knowing positional uncertainties,
and plot the mean value of
the estimators of $f$ listed in Sect.~\ref{failure} 
as a function of $n$ in Fig.~\ref{fig2}.
This time, however, we replaced $\hat f_\oto^\wrong$ by the estimator
$\hat f_\oto$ computed from the $p_\oto$.

For several-to-one simulations, $\hat f_\sto$ is by far the best estimator of $f$
and does not show any \vadjust{\vspace{-1pt}}significant 
bias, whatever the value of $n$.
Estimators $\hat f_\oto$ and $\hat f_\ots$ do not recover the input value
of $f$, which is not \vadjust{\vspace{-1pt}}surprising since they are not built from the right
assumption here;
moreover, while $\hat f_\sto$, $\hat f'_\ots$, and $\hat f_\oto$ are 
obtained by maximizing $\Lhsto$, $\Lhots$, and $\Lhoto$, respectively,
$\hat f_\ots$ is not directly \emph{fitted} to the data.

For one-to-one simulations, and unlike $\hat f_\oto^\wrong$,
$\hat f_\oto$ is a consistent estimator of $f$, as expected.
Puzzlingly, $\hat f_\sto$ also works very well, maybe because \ref{H_sto}
is a more relaxed assumption than~\ref{H_oto};
whatever the reason, this is not a problem.
\subsection{Simultaneous estimation of \texorpdfstring{$f$}{\$f\$} and \texorpdfstring{$\sigmatot$}{\$\textbackslash mathring\textbackslash sigma\$}}
%
\label{simul_f_sigma}
\subsubsection{Circular positional uncertainty ellipses}
\label{circular}
\begin{figure}
  \resizebox{\hsize}{!}{\includegraphics{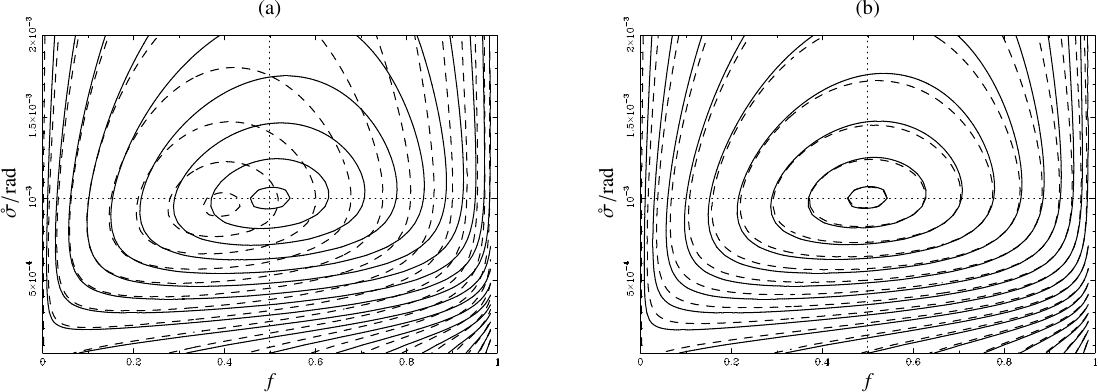}}
  \caption{%
    Contour lines of $\Lhsto$ (solid) and $\Lhoto$
    (dashed) in the $(f, \sigmatot)$ plane.
    Input parameters are the same as in Fig.~\ref{fig2}, except
    that $n = \np = 2\times10^4$;
    the input values of $f$ and $\sigmatot$ are indicated by dotted lines
    (see Sect.~\ref{circular} for details).%
    \enskip \textbf{(a)}~Several-to-one simulations.%
    \enskip \textbf{(b)}~One-to-one simulations.
  }
  \label{fig3}
\end{figure}

How do different estimators of $f$ and $\sigmatot$ behave
when the true values of positional uncertainties are also ignored?
We show in Fig.~\ref{fig3} the result of simulations with the same input as 
in Sect.~\ref{simul_f}, except that $n = \np = 2\times10^4$.
The likelihood $\Lhsto$ peaks very close to the input value of
$\vec x \coloneqq (f, \sigmatot)$ for both types of simulations:
$\hat{\vec x}_\sto$ is still an unbiased estimator of $\vec x$.
For one-to-one simulations, $\Lhoto$ is also maximal near the input value
of $\vec x$, so $\hat{\vec x}_\oto$ is unbiased, too.
\subsubsection{Elongated positional uncertainty ellipses}
\label{elongated}
\begin{figure}
  \resizebox{\hsize}{!}{\includegraphics{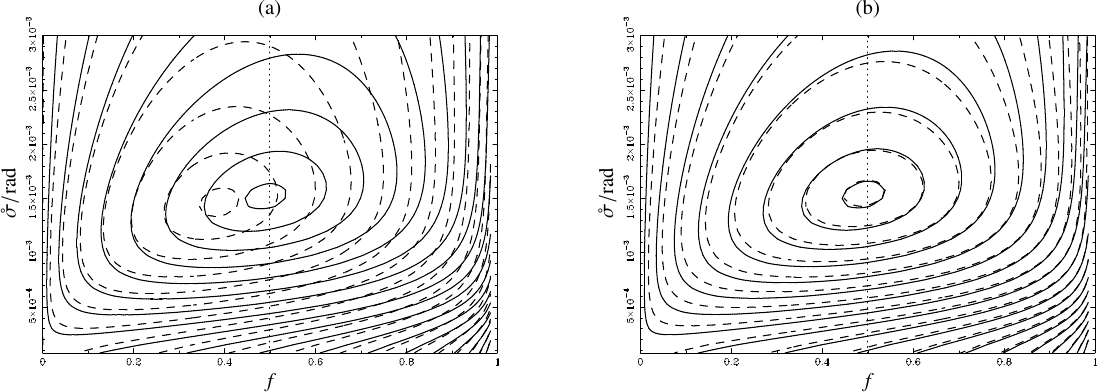}}
  \caption{%
    Contour lines of $\Lhsto$ (solid) and $\Lhoto$
    (dashed) in the $(f, \sigmatot)$ plane.
    Input parameters are the same as in Fig.~\ref{fig2}, except
    that positional uncertainty ellipses are elongated and randomly oriented
    (see Sect.~\ref{elongated} for details);
    the input value of $f$ is indicated by a dotted line.%
    \enskip \textbf{(a)}~Several-to-one simulations.%
    \enskip \textbf{(b)}~One-to-one simulations.
  }
  \label{fig4}
\end{figure}

To test the robustness of estimators of $f$, we ran simulations
with the same parameters, but with elongated positional uncertainty ellipses:
we took $a_i = a'_{\smash[t]{j}} = 1.5\times10^{-3}\,\radian$ and
$b_i = b'_{\smash[t]{j}} = a_i/3$ for all $(M_i, \Mp_j) \in K \times K'$.
These ellipses were randomly oriented; i.e., position angles
(cf.~App.~\ref{identical,known}) $\beta_i$ and $\betapj$ have uniform
random values in $[0, \piup\mathclose{[}$.
We then estimated $f$, but ignoring these positional uncertainties
(see Fig.~\ref{fig4}).

Although the model from which the parameters are fitted is inaccurate here
(the $\xi_{i\comma j}$ are computed assuming \vadjust{\vspace{-1.5pt}}circular positional uncertainties
instead of the unknown elliptical ones), the input value of $f$ is still 
recovered by $\hat f_\sto$ for \vadjust{\vspace{-1pt}}both types of simulations and by $\hat f_\oto$ 
for one-to-one simulations.
The fitting also provides the typical positional uncertainty $\sigmatot$
on the relative positions of associated sources.
\subsection{Choice of association model}
\begin{figure}
  \resizebox{\hsize}{!}{\includegraphics{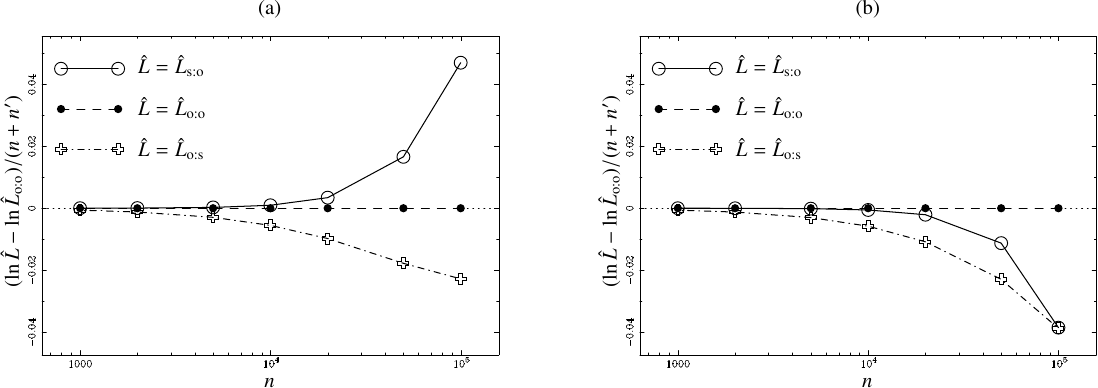}}
  \caption{%
    Normalized average maximum value $\hat\Lh$ of different likelihoods as a 
    function of $n$, compared to $\expandafter\hat\Lhoto$.
    Simulations are the same as in Fig.~\ref{fig2}.%
    \enskip \textbf{(a)}~Several-to-one simulations.%
    \enskip \textbf{(b)}~One-to-one simulations.
  }
  \label{fig5}
\end{figure}

Now, given the two catalogs, which assumption should we adopt
to compute the probabilities $\Prob(A_{i\comma j} \mid C \cap C')$:
several-to-one, one-to-one or one-to-several?
As shown in Fig.~\ref{fig5}, for known positional uncertainties and a given 
$\np$, source confusion is rare at low values of $n$ (there is typically at 
most one possible counterpart) and all assumptions are equally likely.
At larger $n$,
$\expandafter\hat\Lhsto > \expandafter\hat\Lhoto > \expandafter\hat\Lhots$ 
for several-to-one simulations;
as expected, for one-to-one simulations, 
$\expandafter\hat\Lhoto > \expandafter\hat\Lhsto$ and
$\expandafter\hat\Lhoto > \expandafter\hat\Lhots$, with 
$\expandafter\hat\Lhsto \approx \expandafter\hat\Lhots$ for $n = \np$.
In all cases, on average, the right assumption is the most likely.
This is also true when positional uncertainties are ignored
(Sect.~\ref{simul_f_sigma}).

The calculation of $\Lhoto$ is lengthy, and as a substitute to the comparison 
of the likelihoods, the following procedure may be applied to select 
the most appropriate assumption to compute the probabilities of association:
if $\hat f_\sto\multspace n \approx \hat f'_\ots\multspace \np$, use \ref{H_oto};
if $\hat f_\sto\multspace n \not\approx \hat f'_\ots\multspace \np$, then
use \ref{H_sto} if $\hat f_\sto\multspace n > \hat f'_\ots\multspace \np$,
and \ref{H_ots} otherwise.
\section{Conclusion}
In this paper, we computed the probabilities of positional association
of sources between two catalogs $K$ and $K'$ under two different assumptions:
first, the easy case where several $K$-objects may share the same counterpart 
in $K'$,
then the more natural but numerically intensive case of one-to-one 
associations only between $K$ and $K'$.

These probabilities depend on at least one unknown parameter:
the fraction of sources with a counterpart.
If the positional uncertainties are unknown, other parameters are required
to compute the probabilities.
We calculated the likelihood of observing all the $K$-~and $K'$-sources at their
effective positions under each of the two assumptions described above,
and estimated the unknown parameters by maximizing these likelihoods.
The latter are also used to select the best association model.

These relations were implemented in a code, \Aspects, which we make public
and with which we analyzed all-sky several-to-one and one-to-one simulations.
In all cases, the assumption with the highest likelihood is the right one,
and estimators of unknown parameters obtained for it do not show any bias.

In the simulations, we assumed that the density of $K$-~and $K'$-sources
was uniform on the sky area $\Stot$:
the quantities $\xi_{i\comma 0}$ and $\xi_{0\comma j}$ used to compute the 
probabilities are then equal to $1/\Stot$.
If the density of objects is not uniform, we might take
$\xi_{i\comma 0} = \rho(M_i)/n$ and $\xi_{0\comma j} = \rho'\mkern-1mu(\Mp_j)/\np$,
where $\rho$ and $\rho'$ are, respectively, the local surface
densities of $K$-~and $K'$-sources;
but if the $\rho'\!/\rho$ ratio varies on the sky, so will the fraction of 
sources with a counterpart --~something we did not try to model.
Considering clustering or the side effects%
\footnote{%
  The impact of clustering or of side effects on estimators of unknown
  parameters might however easily be tested through simulations.
}
due to a small $\Stot$, as well as taking priors
on the SED of objects into account was also beyond the scope of this paper.

In spite of these limitations, \Aspects\ is a robust tool that should help
astronomers cross-identify astrophysical sources automatically,
efficiently and reliably.
\appendix
\section{Probability distribution of the observed
  relative positions of associated sources}
\label{cov}
\subsection{Properties of normal laws}
\label{Gaussian}
We first recall a few standard results.
The probability that an $m$-dimensional normally distributed random vector
$\vec W$ of mean $\vec\mu$ and variance $\Gamma$ falls in some domain 
$\Omega$ is
\begin{equation}
  \Prob(\vec W \in \Omega) 
  = 
  \int_{\vec w \in \Omega}
  \frac{
    \exp\Bigl(-\frac{1}{2}\multspace \transpose{[\vec w-\vec\mu]_B}
    \cdot \Gamma_B^{-1} \cdot [\vec w-\vec\mu]_B\Bigr)
  }{
    (2\multspace \piup)^{m/2} \multspace \!\sqrt{\det\Gamma_B}
  }
  \multspace \df^m\vec w_B,
\end{equation}
where 
$B \coloneqq (\vec u_1, \ldots, \vec u_m)$ is a basis,
$\vec w$ is a vector, $\vec w_B = \transpose{(w_1, \dotsc, w_m)}$ (resp.\ $\vec\mu_B$) is the
column vector expression of $\vec w$ (resp.\ $\vec\mu$) in $B$, $\df^m\vec w_B \coloneqq \prod_{i=1}^m \df w_i$,
and $\Gamma_B$ is the covariance matrix of $W$ (i.e.\ the matrix representation of $\Gamma$) in $B$.
We denote this by $\vec W \sim G_m(\vec\mu, \Gamma)$.

In another basis
$B' \coloneqq (\vec u'_{\smash[t]{1}}, \ldots, \vec u'_{\smash[t]{m}})$,
we have $\vec w_B = T_{B\rightarrow B'} \cdot \vec w_{B'}$, where
$T_{B\rightarrow B'}$ is the transformation matrix from $B$ to $B'$ (i.e.\
$\vec u'_{\smash[t]{j}} =
\sum_{i=1}^m {(T_{B\rightarrow B'})_{i\comma j}\multspace \vec u_i}$).
Since
$\df^m\vec w_B = \lvert\det T_{B\rightarrow B'}\rvert\multspace \df^m\vec w_{B'}$ and
\begin{equation}
  \transpose{(\vec w-\vec\mu)_B} \cdot \Gamma_B^{-1} \cdot (\vec w-\vec\mu)_B 
  = 
  \transpose{(\vec w-\vec\mu)_{B'}} \cdot \Bigl(T_{B\rightarrow B'}^{-1} \cdot
  \Gamma_B \cdot \transpose{[T_{B\rightarrow B'}^{-1}]}\Bigr)^{-1} \cdot 
  (\vec w-\vec\mu)_{B'},
\end{equation}
we still obtain
\begin{equation}
  \Prob(\vec W \in \Omega) 
  = 
  \int_{\vec w \in \Omega}
  \frac{
    \exp\Bigl(-\frac{1}{2}\multspace \transpose{[\vec w-\vec\mu]_{B'}}
    \cdot \Gamma_{B'}^{-1} \cdot [\vec w-\vec\mu]_{B'}\Bigr)
  }{
    (2\multspace \piup)^{m/2} \multspace \!\sqrt{\det\Gamma_{B'}}
  }
  \multspace \df^m\vec w_{B'},
\end{equation}
where 
$\Gamma_{B'} \coloneqq T_{B\rightarrow B'}^{-1} \cdot \Gamma_B \cdot
\transpose{(T_{B\rightarrow B'}^{-1})}$
is the covariance matrix of $\vec W$ in $B'$.
In the following, $B$ and $B'$ are orthonormal bases, so $T_{B\rightarrow B'}$ 
is a rotation matrix.
From $\transpose{T_{B\rightarrow B'}} = T_{B\rightarrow B'}^{-1}$, we get
\begin{equation}
  \Gamma_{B'} = 
  \transpose{T_{B\rightarrow B'}} \cdot \Gamma_B \cdot T_{B\rightarrow B'}.
\end{equation}

For independent random vectors 
$\vec W_1 \sim G_m(\vec\mu_1, \Gamma_1)$ and
$\vec W_2 \sim G_m(\vec\mu_2, \Gamma_2)$, we have
\begin{equation}
  \label{somme_gaussiennes}
  \vec W_1 \pm \vec W_2 \sim G_m(\vec\mu_1 \pm \vec\mu_2, \Gamma_1+\Gamma_2).
\end{equation}
\subsection{Covariance matrix of the probability distribution
  of relative positions}
\label{cov_mat}
We now use these results to derive the probability distribution of vector
$\vec r_{i\comma j} \coloneqq \vrpj-\vec r_i$, where $\vec r_i$ and $\vrpj$
are, respectively, the observed positions of source $M_i$ of $K$
and of its counterpart $\Mp_j$ in $K'$.
Introducing the true positions $\vrzi$ and $\vrpzj$ of $M_i$ and $\Mp_j$, 
we have
\begin{equation}
  \label{rel_pos}
  \vec r_{i\comma j} = (\vrpj - \vrpzj) + (\vrpzj - \vrzi) + (\vrzi -\vec r_i).
\end{equation}
\subsubsection{Covariance matrix for identical true positions
  and known positional uncertainties}
\label{identical,known}
Assume%
\footnote{%
  None of the results established outside of App.~\ref{cov} depends on this 
  assumption.
},
as is usual, that
\begin{equation}
  \label{r_i,r'_j}
  \vec r_i - \vrzi \sim G_2(\vec 0,\Gamma_i)
  \qquad\text{and}\qquad
  \vrpj - \vrpzj \sim G_2(\vec 0,\Gammapj).
\end{equation}
If the true positions of $M_i$ and $\Mp_j$ are identical (case of point 
sources), then, 
from Eqs.~\eqref{somme_gaussiennes}, \eqref{rel_pos}, and \eqref{r_i,r'_j},
\begin{equation}
  \label{rel_pos_ident}
  \vec r_{i\comma j} \sim G_2(\vec 0, \Gamma_{i\comma j}),
  \quad\text{where }
  \Gamma_{i\comma j} \coloneqq \Gamma_i + \Gammapj.
\end{equation}
(See also \citealt{Condon95}.)
In Eqs.~\eqref{r_i,r'_j}, $\vec r_i - \vrzi$ and $\vrpj - \vrpzj$
must be considered as the \emph{projections} (gnomonic ones, for instance)
of these vectors on the planes tangent to the sphere at $M_i$ and $\Mp_j$, 
respectively;
Eqs.~\eqref{r_i,r'_j} are approximations, \vadjust{\vspace{-2pt}}valid only because positional 
uncertainties are small%
\footnote{%
  If it were not the case, the probability of $\vec r_i - \vrzi$ and 
  $\vrpj - \vrpzj$ might be modeled using \citet{Kent} distributions
  (an adaptation to the sphere of the planar normal law),
  but no result like Eq.~\eqref{rel_pos_ident} would then hold:
  unlike Gaussians, Kent distributions are not \emph{stable}.
}.
Equation~\eqref{rel_pos_ident} is also an approximation:
it is appropriate because the observed positions of associated sources
$M_i$ and $\Mp_j$ are close, so the tangent planes to the sphere at both points
nearly coincide.

To use Eq.~\eqref{rel_pos_ident}, we now compute the column vector expression of 
$\vec r_{i\comma j}$ and the covariance matrices associated to $\Gamma_i$,
$\Gammapj$, and $\Gamma_{i\comma j}$ in some common basis.
For convenience, we drop the subscript and the ``prime'' symbol in the 
following whenever an expression only depends on either $M_i$ or $\Mp_j$.

Let $(\vec u_x, \vec u_y, \vec u_z)$ be a direct orthonormal basis, with
$\vec u_z$ oriented from the Earth's center $O$ to the North Celestial Pole
and $\vec u_x$ from $O$ to the Vernal Point.
At a point $M$ of right ascension $\alpha$ and declination $\delta$,
a direct orthonormal basis $(\vec u_r, \vec u_\alpha, \vec u_\delta)$ is defined by
\begin{align}
  \vec u_r
  &\coloneqq
  \frac{\vec{OM}}{\lVert\vec{OM}\rVert} 
  =
  \cos\delta\multspace \cos\alpha\multspace \vec u_x
  + \cos\delta\multspace \sin\alpha\multspace \vec u_y 
  + \sin\delta\multspace \vec u_z,
  \label{u_r}
  \\
  \vec u_\alpha
  &\coloneqq
  \frac{\partial\vec u_r/\partial\alpha}{
    \lVert\partial\vec u_r/\partial\alpha\rVert}
  =
  -\!\sin\alpha\multspace \vec u_x + \cos\alpha\multspace \vec u_y,
  \label{u_alpha}
  \\
  \vec u_\delta
  &\coloneqq
  \frac{\partial\vec u_r/\partial\delta}{
    \lVert\partial\vec u_r/\partial\delta\rVert}
  =
  -\!\sin\delta\multspace \cos\alpha\multspace \vec u_x 
  - \sin\delta\multspace \sin\alpha\multspace \vec u_y
  + \cos\delta\multspace \vec u_z.
  \label{u_delta}
\end{align}

The uncertainty ellipse on the position of $M$ is characterized by the lengths 
$a$ and $b$ of its semi-major and semi-minor axes, and by the position angle 
$\beta$ between the north and the semi-major axis.
Let $\vec u_a$ and $\vec u_b$ be unit vectors directed along 
the major and the minor axes, respectively,
and such that $(\vec u_r, \vec u_a, \vec u_b)$ 
is a direct orthonormal basis and that
$\beta \coloneqq \angle(\vec u_\delta, \vec u_a)$
is in $[0, \piup\mathclose{[}$ when counted eastward.
Since $(\vec u_\alpha, \vec u_\delta)$ is obtained from $(\vec u_a, \vec u_b)$
by a $(\beta-\piup/2)$-counterclockwise rotation in the plane oriented by
$+\vec u_r$, 
we have
$T_{(\vec u_a{,}\, \vec u_b)\rightarrow(\vec u_\alpha{,}\, \vec u_\delta)} = 
\Rot(\beta-\piup/2)$,
where, for any angle $\tau$,
\begin{equation}
  \Rot\tau 
  \coloneqq
  \Left(\begin{matrix}
    \cos\tau & -\!\sin\tau
    \\
    \sin\tau & \cos\tau
  \end{matrix}\Right).
\end{equation}
Using notation
\begin{equation}
  \Diag\bigl(d_1, d_2\bigr) 
  \coloneqq
  \Left(\begin{matrix}
    d_1 & 0
    \\
    0 & d_2
  \end{matrix}\Right)
\end{equation}
for diagonal matrices, we have%
\footnote{%
  We seize this opportunity to correct equations (A.8) to (A.11)
  of \citet{Pineau}:
  $a$ and $b$ should be replaced by their squares in these formulae.
}
$\Gamma_{(\vec u_a{,}\, \vec u_b)} = \Diag\bigl(a^2, b^2\bigr)$ and
\begin{equation}
  \Gamma_{(\vec u_\alpha{,}\, \vec u_\delta)} =
  \transpose{\Rot}(\beta-\piup/2) \cdot \Diag\bigl(a^2, b^2\bigr)
  \cdot \Rot(\beta-\piup/2).
\end{equation}

As noticed by \citet{Pineau}, around the Poles, even for sources $M_i$ and
$\Mp_j$ close to each other, we may have 
$(\uai, \udi) \not\approx (\uapj, \udpj)$:
the covariance matrices $(\Gamma_i)_{(\uai,\,\udi)}$
and $(\Gammapj)_{(\uapj,\,\udpj)}$ must therefore be first
converted to a common basis before their summation in Eq.~\eqref{rel_pos_ident}.
We use the same basis as \citet{Pineau}, denoted by $(\vec t, \vec n)$ below.
While the results we get are intrinsically the same, some people may find
our expressions more convenient.

Denote by $\vec n \coloneqq \uri \times \urpj/\lVert\uri \times \urpj\rVert$
a unit vector perpendicular to the plane $(O, M_i, \Mp_j)$.
Because $\psi_{i\comma j} \coloneqq \angle(\uri, \urpj) \in [0, \piup]$, 
we have $\uri\cdot\urpj = \cos\psi_{i\comma j}$ and 
$\lVert\uri \times \urpj\rVert = \sin\psi_{i\comma j}$, so
\begin{equation}
  \label{psi_arccos}
  \psi_{i\comma j} = \arccos\Bigl(
  \cos\delta_i\multspace \cos\deltapj\multspace \cos[\alphapj-\alpha_i]
  + \sin\delta_i\multspace \sin\deltapj
  \Bigr),
\end{equation}
and
\begin{equation}
  \label{vec_n}
  \vec n = \frac{\uri \times \urpj}{\sin\psi_{i\comma j}}.
\end{equation}

Let $\gamma_i \coloneqq \angle(\vec n, \udi)$ and
$\gammapj \coloneqq \angle(\vec n, \udpj)$ be angles oriented clockwise around
$+\uri$ and $+\urpj$, respectively.
Angle $\gamma_i$ is fully determined by the following expressions
(cf.~Eqs.~\eqref{vec_n}, \eqref{u_alpha} and \eqref{u_delta}):
\begin{align}
  \cos\gamma_i
  &=
  \vec n\cdot \udi 
  =
  \frac{(\uri \times \urpj) \cdot \udi}{\sin\psi_{i\comma j}}
  =
  \frac{(\udi \times \uri) \cdot \urpj}{\sin\psi_{i\comma j}}
  =
  \frac{\uai\cdot \urpj}{\sin\psi_{i\comma j}}
  =
  \frac{\cos\deltapj\multspace \sin(\alphapj-\alpha_i)}{\sin\psi_{i\comma j}};
  \\
  \sin\gamma_i
  &=
  -\vec n\cdot \uai
  =
  -\frac{(\uri \times \urpj) \cdot \uai}{\sin\psi_{i\comma j}}
  =
  -\frac{(\uai \times \uri) \cdot \urpj}{\sin\psi_{i\comma j}}
  =
  \frac{\udi\cdot \urpj}{\sin\psi_{i\comma j}}
  =
  \frac{\cos\delta_i\multspace \sin\deltapj -
    \sin\delta_i\multspace \cos\deltapj\cos(\alphapj-\alpha_i)}{
    \sin\psi_{i\comma j}}.
\end{align}
Similarly,
\begin{equation}
  \cos\gammapj 
  = 
  \frac{\cos\delta_i\multspace \sin(\alphapj-\alpha_i)}{\sin\psi_{i\comma j}}
  \qquad \text{and} \qquad
  \sin\gammapj 
  = 
  \frac{\cos\delta_i\multspace \sin\deltapj\cos(\alphapj-\alpha_i)
    - \sin\delta_i\multspace \cos\deltapj}{\sin\psi_{i\comma j}}.
\end{equation}

Let
\begin{equation}
  \vec t \coloneqq \vec n \times \uri
\end{equation}
($\approx \vec n \times \urpj$ since $M_i$ and $\Mp_j$ are close):
vector $\vec t$ is a unit vector tangent in $M_i$ to the minor arc of great 
circle going from $M_i$ to $\Mp_j$.
Project the sphere on the plane $(M_i, \vec t, \vec n)$ tangent
to the sphere in $M_i$ (the specific projection does not matter since we
consider only $K'$-sources in the neighborhood of $M_i$).
We have
\begin{equation}
  \vec r_{i\comma j} \approx \psi_{i\comma j}\multspace \vec t,
\end{equation}
and the basis $(\vec t, \vec n)$ is obtained from $(\vec u_a, \vec u_b)$
by a $(\beta+\gamma-\piup/2)$-counterclockwise rotation around $+\vec u_r$, so,
\begin{align}
  (\Gamma_i)_{(\vec t,\,\vec n)}
  &=
  \transpose{\Rot}(\beta_i+\gamma_i-\piup/2) \cdot \Diag\bigl(a_i^2, b_i^2\bigr)
  \cdot \Rot(\beta_i+\gamma_i-\piup/2),
  \\
  (\Gammapj)_{(\vec t,\,\vec n)}
  &=
  \transpose{\Rot}(\betapj+\gammapj-\piup/2) \cdot
  \Diag\bigl(a_{\smash[t]{j}}'^2, b_{\smash[t]{j}}'^2\bigr)
  \cdot \Rot(\betapj+\gammapj-\piup/2).
\end{align}
\subsubsection{Case of unknown positional uncertainties}
\label{unknown}
If the positional uncertainty on $M_i$ is unknown, we may model it with
$(\Gamma_i)_{(\vec t,\,\vec n)} = \sigma^2\multspace \Diag(1, 1)$,
using the same $\sigma$
for all $K$-sources, and derive an estimate of $\sigmatot \coloneqq \sigma$ by 
maximizing the likelihood to observe the distribution of $K$-~and $K'$-sources
(see Sects.~\ref{fractionsto} and \ref{fractionoto}).
For a galaxy, however, the positional uncertainty on its center is likely to 
increase with its size.
If the position angle $\theta_i$ (counted eastward from the north) and the 
major and minor diameters $D_i$ and $d_i$ of the best-fitting ellipse of some 
isophote are known for $M_i$ (for instance, parameters $\textnormal{PA}$, 
$D_{25}$ and $d_{25} \coloneqq D_{25}/R_{25}$ taken from the \textsc{RC}3 catalog
\citep{RC3} or \textsc{HyperLeda} \citep{HYPERLEDA}), we may model the 
positional uncertainty with
\begin{align}
  (\Gamma_i)_{(\vec t,\,\vec n)}
  &=
  \transpose{\Rot}(\theta_i+\gamma_i-\piup/2) \cdot
  \Diag\Bigl(\sigma^2 + [\nu\multspace D_i]^2, 
  \sigma^2 + [\nu\multspace d_i]^2\Bigr)
  \cdot \Rot(\theta_i+\gamma_i-\piup/2)
  \notag \\
  &=
  \sigma^2\multspace \Diag(1, 1) 
  + \nu^2\multspace \transpose{\Rot}(\theta_i+\gamma_i-\piup/2)
  \cdot \Diag\bigl(D_i^2, d_i^2\bigr) \cdot \Rot(\theta_i+\gamma_i-\piup/2),
\end{align}
and derive estimates of $\sigmatot \coloneqq \sigma$ and $\nutot \coloneqq \nu$
from the likelihood.
Such a technique might indeed be used to estimate the accuracy of coordinates
in some catalog (see \citet{PP} for another method).

If the positional uncertainty on $\Mp_j$ is unknown too, we can also put
\begin{equation}
  (\Gammapj)_{(\vec t,\,\vec n)}
  =
  \sigma'^2\multspace \Diag(1, 1) 
  + \nu'^2\multspace \transpose{\Rot}(\thetapj+\gammapj-\piup/2) \cdot
  \Diag\bigl(D_i^2, d_i^2\bigr) \cdot \Rot(\thetapj+\gammapj-\piup/2),
\end{equation}
with the same $\sigma'$ and $\nu'$ for all $K'$-sources.
As $\gammapj + \thetapj = \gamma_i + \theta_i$,
\vadjust{\vspace{-2pt}}only estimates of $\sigmatot \coloneqq \bigl(\sigma^2+\sigma'^2\bigr)^{1/2}$ and
$\nutot \coloneqq \bigl(\nu^2+\nu'^2\bigr)^{1/2}$ may be obtained%
\footnote{%
  However, as noticed by \citet{dVH} in a different context, if three samples 
  with unknown uncertainties $\sigma_i$ ($i \in \integinterv{1}{3}$) are 
  available and if the combined uncertainties
  $\sigma_{i\comma j} \coloneqq (\sigma_i^2+\sigma^2_j)^{1/2}$
  may be estimated for all the pairs 
  $(i, j)_{j\neq i} \in \integinterv{1}{3}^2$,
  as in our case, then $\sigma_i$ may be determined for each sample.
  \citet{PP} used this technique to compute the accuracy of galaxy coordinates.
}
by maximizing the likelihood, not the values of $\sigma$, $\sigma'$, $\nu$ or 
$\nu'$ themselves.
\subsubsection{Possibly different true positions}
\label{different}
A similar technique can be applied if the true centers of $K$-sources
and of their counterparts in $K'$ sometimes differ.
This might be useful in particular when associating galaxies from an
optical catalog and from a ultraviolet or far-infrared one, because, while
the optical is dominated by smoothly-distributed evolved stellar populations,
the ultraviolet and the far-infrared mainly trace star-forming regions.
Observations of galaxies (e.g., \citealt{Kuchinski}) have indeed shown that 
galaxies are very patchy in the ultraviolet, and the same has been observed 
in the far-infrared.

Since the angular distance between the true centers should increase with the
size of the galaxy, we might model this as
\begin{equation}
  \vrpzj-\vrzi\sim G_2(\vec 0, \Gammazi),
  \quad\text{where }
  (\Gammazi)_{(\vec t,\,\vec n)}
  = \nu_0^2\multspace
  \transpose{\Rot}(\theta_i+\gamma_i-\piup/2) \cdot \Diag\bigl(D_i^2, d_i^2\bigr)
  \cdot \Rot(\theta_i+\gamma_i-\piup/2).
\end{equation}
We then have
\begin{equation}
  \vec r_{i\comma j}
  \sim G_2(\vec 0, \Gamma_{i\comma j}),
  \quad \text{with }
  \Gamma_{i\comma j} \coloneqq \Gamma_i+\Gammapj+\Gammazi.
\end{equation}
Once again, if $\sigma$, $\sigma'$, $\nu$, $\nu'$ and $\nu_0$ are unknown,
only $\sigmatot \coloneqq \bigl(\sigma^2+\sigma'^2\bigr)^{1/2}$ and
$\nutot \coloneqq \bigl(\nu^2+\nu'^2+\nu_0^2\bigr)^{1/2}$ may be estimated
through likelihood maximization.
\begin{acknowledgements}
  The initial phase of this work took place at the NASA/Goddard Space Flight 
  Center, under the supervision of Eli Dwek, and was supported by the National 
  Research Council through the Resident Research Associateship Program.
  We acknowledge them sincerely.
  We also thank St\'ephane Colombi for the discussions we had on the 
  properties of maximum likelihood estimators.
\end{acknowledgements}
\bibliographystyle{aa}
\bibliography{references}
\end{document}